\documentclass[review]{elsarticle}

\journal{Journal of \LaTeX\ Templates}

\def\ind{\begin{picture}(9,8)
         \put(0,0){\line(1,0){9}}
         \put(3,0){\line(0,1){8}}
         \put(6,0){\line(0,1){8}}
         \end{picture}
        }

    \usepackage{multirow}
\usepackage{subfigure}
\usepackage{amsbsy,amsthm}
\usepackage{amssymb,amsmath}
\usepackage{colortbl}
    \usepackage{bm}
    \usepackage{graphicx,color}
    \usepackage{graphicx,psfrag,epsf}
    
    \theoremstyle{definition}

\newtheorem{example}{Example}

\usepackage{numcompress}
    
    \newcommand{\sumn}{\sum_{i=1}^n}

\begin{document}

\begin{frontmatter}

\title{Robust Modeling Using Non-Elliptically Contoured Multivariate $t$ Distributions}

\author{Zhichao Jiang}
\address{School of Mathematical Sciences, Peking University,  Bejing 100871, China}

\author{Peng Ding}
\address{
Department of Statistics, University of California, Berkeley, California 94720, USA}

\cortext[mycorrespondingauthor]{Corresponding author}
\ead{pengdingpku@berkeley.edu}


\begin{abstract}
Models based on multivariate $t$ distributions are widely applied to analyze data with heavy tails. However, all the marginal distributions of the multivariate $t$ distributions are restricted to have the same degrees of freedom,  making these models unable to  describe different marginal heavy-tailedness.   We generalize the traditional multivariate $t$ distributions to  non-elliptically contoured multivariate $t$ distributions, allowing for different marginal degrees of freedom. We apply the non-elliptically contoured multivariate $t$ distributions to three widely-used models: the Heckman selection model with different degrees of freedom for selection and outcome equations, the multivariate Robit model with different degrees of freedom for  marginal responses, and the linear mixed-effects model with different degrees of freedom for random effects and within-subject errors. Based on the Normal mixture representation of our $t$ distribution, we propose efficient Bayesian inferential procedures for the model parameters based on data augmentation and parameter expansion. We show via simulation studies and  real examples  that the conclusions are sensitive  to the existence of different marginal heavy-tailedness.
\end{abstract}


\end{frontmatter}

%\linenumbers

\section{Introduction}
\label{sec:intro}

Normal distributions are widely used for statistical modeling due to their simplicity and interpretability. Many results and methods, such as ordinary least squares, can be derived analytically when the relevant variables are Normally distributed.  However, in practice, data may have heavy tails, which are difficult to deal with using Normal models.

Models based on $t$ distributions are frequently applied for robust analysis \citep{zellner1976bayesian,lange1989robust, geweke1994priors, liu1999parameter, liu2004robit, gelman2014bayesian, zhang2014robust}, and they are attractive generalizations of the models based on Normal distributions such as linear and Probit models. \citet{student1908probable} proposes the classical univariate $t$ distribution, which is symmetric and bell-shaped, but has heavier tails than the standard Normal distribution. A multivariate  $t$ distribution (MTD) is a multivariate generalization of the one-dimensional Student $t$ distribution. Because it is elliptically contoured, any linear transformations follow $t$ distributions with the same number of degrees of freedom.  However, it is sometimes too restrictive to require all marginal degrees of freedom be the same. Previous literature generalizes the MTD through different ways. For a recent review, see Nadarajah and Dey's paper \cite{nadarajah2005multitude}. \citet{arellano1995some} discuss  three characterizations of  the MTD within the class of elliptical contoured distributions \citep{cambanis1981theory}. \citet{fang2002meta} propose the meta-elliptical distributions using copula. \citet{jones2002dependent} develops a dependent bivariate $t$ distribution with different marginal degrees of freedom. However, none of their work allows the marginal distributions to be independent, which is a limitation for modeling. In this paper, we propose a non-elliptically contoured multivariate $t$ distribution (NECTD), allowing for different marginal degrees of freedom and independent  marginal distributions. The bivariate case of the NECTD is similar to the formulation of \citet{shaw2008bivariate}. Our NECTD, based on scale mixtures of the components of the multivariate Normal distribution, are flexible enough to be incorporated into various models and enjoy easy Bayesian computation using data augmentation \citep{tanner1987calculation} and parameter expansion \citep{liu1998parameter,meng1999seeking,van2001art}. We further illustrate its potential applications by generalizing the Heckman selection model, multivariate Robit model, and linear mixed-effects model.

Sample selection \citep{heckman1979sample} or missing data \citep{little2002statistical} problems are common in applied research. The Heckman selection model \citep{heckman1979sample} is the most famous model dealing with sample selection, which consists of a Probit selection equation and a linear outcome equation. To deal with heavy-tailed data with sample selection,  \citet{marchenko2012aheckman} propose a Heckman selection-$t$ model, modeling the error terms of the selection and outcome equations as a bivariate $t$ distribution.  However, in the Heckman selection-$t$ model, the error terms  are constrained to have the same number of degrees of freedom, which cannot handle cases with different heavy-tailedness in the selection  and outcome equations.  Ignoring the heterogeneity of the marginal numbers of degrees of freedom may lead to biased inference. In order to overcome this limitation, we propose a generalized selection-$t$ model based on the NECTD, allowing for different heavy-tailedness in the selection and outcome equations.

The Logistic or Probit model for binary data can be represented by a latent linear model with a Logistic or Normal error distribution \citep{albert1993bayesian}. To make such commonly-used models more robust to outliers, \citet{liu2004robit} proposes a Robit regression model, replacing the error in the latent linear model by a $t$ distribution. When generalizing the Robit model to multivariate settings, it may be restrictive to have all the marginal distributions sharing the same number of degrees of freedom. Fortunately, we can generalize the multivariate Robit model by assuming NECTD error terms.

The linear mixed-effects model is frequently used for analyzing repeatedly measured data \citep{hartley1967maximum, laird1982random}. It assumes Normal distributions for both the random effects and the within-subject errors. \citet{pinheiro2001efficient} propose a robust linear mixed-effects model, in which the random effects and the within-subject errors follow a MTD.  This model is widely used in practice \citep{lin2006robust, lin2007bayesian}. However, their model restricts the numbers of degrees of freedom of the random effects and the within-subject errors to be the same. Based on the NECTD, we propose a generalized linear $t$ mixed-effects model, allowing for different heavy-tailedness in the two sources of variations.

The paper proceeds as follows. We introduce the NECTD and discuss its statistical properties in Section 2. In Sections 3--5, we propose the generalized selection-$t$, Robit, and linear $t$ mixed-effects models, respectively. For each model, we propose a Bayesian inferential procedure for the parameters, give a numerical example, and show its application on a real dataset. We conclude with a discussion in Section 6. In Appendices A and B, we present the properties of the NECTD and provide the details of Bayesian inference for NECTD.
In Appendices C, D and E, we provide the details for Bayesian posterior computation. In Appendix F, we provide the  sensitivity analysis for our three real data examples.

\section{Non-Elliptically Contoured Multivariate $t$ Distribution}
\label{sec:meth}
The traditional $p$-dimensional MTD, $\bm{t}_p(\bm{\mu}, \bm{\Sigma}, \nu)$,
has probability density function:
\begin{eqnarray}
f(\bm{x}) =
\frac{\Gamma\left(\frac{ \nu+p}{2} \right) }{\Gamma\left( \frac{\nu}{2} \right)\nu^{p/2}\pi^{p/2}\left|{\boldsymbol\Sigma}\right|^{1/2}}
\left\{ 1+ \nu^{-1}(\bm{x}-{\boldsymbol\mu})^\top{\boldsymbol\Sigma}^{-1}(\bm{x}-{\boldsymbol\mu})\right\} ^{-(\nu+p)/2},
\end{eqnarray}
where $\bm{\mu}$ is the location parameter, $\bm{\Sigma}$ is the scale matrix, and $\nu$ is the number of degrees of freedom.

Let $\bm{I}_{p}$ denote a $p\times p$ identity matrix. We can represent the MTD  as a ratio between a multivariate Normal random vector and the square root of an independent Gamma random variable:
$$
\bm{X} \mid q  \sim \bm{N}_p(\bm{\mu}, \bm{\Sigma} / q), \quad  q \sim \chi^2_\nu/\nu,
$$
or equivalently,
\begin{eqnarray}
\bm{X}  = \bm{\mu} +  q^{-1/2}\bm{\Sigma}^{1/2} \bm{Z}, \quad \bm{Z}\sim \bm{N}_p(\bm{0}, \bm{I}_p), \quad q \sim \chi^2_\nu/\nu, \quad  q\ind \bm{Z}. \label{eqn:trep}
\end{eqnarray}
The additional factor $q$ with $E(q)=1$ does not change the location but amplifies the variability of the multivariate Normal distribution $\bm{N}_p(\bm{\mu}, \bm{\Sigma} )$. When $q$ falls close to zero, the MTD produces extreme values. 
Representation (\ref{eqn:trep}) implies that each marginal distribution of $\bm{X}$ follows a univariate $t$ distribution with the same number of degrees of freedom $\nu$, namely, $X_{j}\sim t_1(\mu_j, \sigma_j^2, \nu)$. Moreover, the traditional MTD is an elliptically contoured distribution, which enjoys nice mathematical properties \citep{fang1990symmetric,anderson2003introduction, kotz2004multivariate}.

However, the constraint of a common number of degrees of freedom prevents modeling multivariate data with different heavy-tailedness in different dimensions. We tackle this problem by generalizing the traditional elliptically contoured MTD. Let $\bm{Q}  = \text{diag}  \{ q_{1}\bm{I}_{p_1}, \ldots, q_{s}\bm{I}_{p_s} \} $ be a block diagonal matrix with $\sum_{j=1}^{s}p_j=p$ and $ \{ q_{j}\sim \chi^2_{\nu_j}/\nu_j: j=1, \ldots, s\}$.
Instead of using the probability density function, we define NECTD using a scale mixture of a Normal random vector:
\begin{eqnarray}
\bm{X}  =  \bm{\mu} +  \bm{Q}^{-1/2}   \bm{\Sigma}^{1/2}  \bm{Z}, \quad \bm{Z}\sim \bm{N}_p(\bm{0}, \bm{I}_p), \label{def:nectd}
\end{eqnarray}
where  $\{\bm{Z}, q_{j}: j=1, \ldots, s\}$ are mutually independent. Let $\bm{t}_p(\bm{\mu}, \bm{\Sigma}, \bm{p}, \bm{\nu})$ denote an NECTD, where $\bm{\nu} = (\nu_1, \ldots, \nu_s)^\top$ and $\bm{p} = (p_1, \ldots, p_s)^\top$.

Marginally, for $s_1=\sum_{j=1}^{m-1}p_j+1$ and $s_2=\sum_{j=1}^{m}p_j$, we have $(X_{s_1},\ldots, X_{s_2})\sim \bm{t}_{p_m}(\bm{\mu}_m, \bm{\Sigma}_m, \nu_m)$, where $\bm{\mu}_m $ and $\bm{\Sigma}_m$ are the corresponding location vector and scale matrix of $(X_{s_1},\ldots, X_{s_2})$. 
Therefore, our NECTD is a generalization of the traditional MTD.

An alternative way to generalize the multivariate $t$ distribution is through linear transformations of $t$ random variables with different numbers of degrees of freedom. This is equivalent to swapping $\bm{Q}$ and $\bm{\Sigma}$  in \eqref{def:nectd}. When $p_1=\cdots=p_s$,  it is the independent component model proposed by \citet{ilmonen2011semiparametrically}. However, under this model, the distribution does not have marginal $t$ distributions.

The NECTD has many properties similar to the MTD. For example, each component of an NECTD follows a univariate $t$ distribution. However, unlike the MTD, the NECTD is not an elliptically contoured distribution, and thus its linear transformations may not follow $t$ distributions. Generally, the density of the NECTD is very complicated. But we can obtain its density when $p=2$. We present the moments and density of the NECTD in Appendix A.

 An example below further shows the differences between the MTD and NECTD.
 
% \begin{example}
% \label{ex:1}
%Suppose $(Y_1,Y_2)$ follows an NECTD, i.e., $Y_1 =X_1/q_1$ and $Y_2 =X_2/q_2$, with 
%$$
%\begin{pmatrix}
%X_1\\
%X_2
%\end{pmatrix}
%\sim \bm{N}_{2}\left\{ \begin{pmatrix}
%0\\
%0
%\end{pmatrix}, \begin{pmatrix}
%1 & \rho \\
%\rho& 1
%\end{pmatrix}\right\},
%$$
%and $(q_1,q_2, X_1, X_2)$ are mutually independent. We can easily show that $\text{Cov}(Y_1,Y_2)=0$. However, if $(Y_1',Y_2')$ follows a MTD, i.e., $Y_1' =X_1/q_1$ and $Y_2 '=X_2/q_1$, then we have
%\begin{eqnarray*}
%\text{Cov}(Y_1',Y_2')=E\{ \text{Cov}(Y_1',Y_2' \mid q_1)\} +\text{Cov}\{E(Y_1' \mid q_1),E(Y_2'\mid q_1)\}= \text{Var}(1/q_1) \neq 0.
%\end{eqnarray*}
%Thus, even $X_1$ and $X_2$ are independent, the common denominator makes $Y_1'$ and $Y_2'$ correlated. 
%\end{example}
%
%
%Example \ref{ex:1} shows that  the MTD may induce spurious correlation for its marginal components, if both $Y_1$ and $Y_2$ have non-zero means. The following example shows another intriguing difference between the MTD and the NECTD.

\begin{example}
\label{ex:2}
 Suppose $(Y_1,Y_2)$ follows an NECTD, i.e., $Y_1 =Z_1/q_1$ and $Y_2 =Z_2/q_2$, with  
$$
\begin{pmatrix}
Z_1\\
Z_2
\end{pmatrix}
\sim \bm{N}_{2}\left\{ \begin{pmatrix}
0\\
0
\end{pmatrix}, \begin{pmatrix}
1 & \rho \\
\rho& 1
\end{pmatrix}\right\} . 
$$
Suppose $(Y'_1,Y'_2)$ follows a bivariate $t$ distribution, i.e., $Y'_1 =Z_1/q_1$ and $Y'_2 =Z_2/q_1$. If $q_1$ and $q_2$ follow scaled chi-squared distributions with the same degrees of freedom, then
\begin{eqnarray*}
\text{Cov}(Y'_1,Y'_2)&=&E\{ \text{Cov}(Y'_1,Y'_2 \mid q_1)\} +\text{Cov}\{E(Y'_1 \mid q_1),E(Y'_2\mid q_1)\}\\
&=& \rho E(1/q^2_1),\\
\text{Cov}(Y_1,Y_2)&=&E\{ \text{Cov}(Y_1,Y_2 \mid q_1,q_2)\} +\text{Cov}\{E(Y_1 \mid q_1),E(Y_2\mid q_2)\}\\
&=& \rho E(1/q_1)E(1/q_2)
= \rho E^2(1/q_1),
\end{eqnarray*} 
implying $\text{Cov}(Y_1,Y_2) \leq \text{Cov}(Y'_1,Y'_2).$

If $\rho=0$, then $\text{Cov}(Y_1,Y_2) = \text{Cov}(Y'_1,Y'_2) = 0.$ The NECTD has independent components, but the MTD has dependent components.
\end{example}

Therefore, even if the data do have the same marginal degrees of freedom, the correlation structure under NECTD-based models differ from that under MTD-based models. For the same data, the estimated $\bm{\Sigma}$ in \eqref{eqn:trep} and \eqref{def:nectd} may be different. Moreover, the NECTD can handle the case with independent components while the MTD cannot.

We propose a Bayesian inferential procedure for the parameters of the NECTD using the Markov chain Monte Carlo (MCMC). Because inference for the NECTD is a special case of the later models, we present all the details in  Appendix B.

To illustrate the potential applications of the new NECTD in robust data analysis, we will use it to generalize three widely-used models
in the following three sections.

\section{Generalized Selection-$t$ Model}
\label{sec:verify}
\subsection{Model}

Sample selection or missing data is common in applied research. To deal with sample selection,
\citet{heckman1979sample} proposes the Heckman selection model, aiming to  estimate the wage offer function of women. Because housewives' wages are not observed, the sample collected is subject to the self-selection problem. The Heckman selection model consists of a linear equation for the outcome, and a Probit equation for the sample selection mechanism. The outcome equation is
$$
y_i^* = \bm{x}_i^\top \bm{\beta} + \varepsilon_i,
$$
and the sample selection mechanism is characterized by the following latent linear equation:
$$
u_i^* = \bm{w}_i^\top \bm{\gamma} + \eta_i,
$$
for $i=1,\ldots,N.$ 
The indicator for sample selection is $u_i = I(u^*_i>0)$. Let $y_i$ be the observed outcome. We observe the outcome $y^*_i$ if and only if $u^*_i>0$, i.e., $y_i=y^*_i$ if $u_i=1$, and $y_i=NA$ if $u_i=0$, where ``$NA$'' indicates missing data.

Let $K$ and $L$ denote the dimensions of $\bm{x}_i$ and $\bm{w}_i$, respectively. Heckman
\cite{heckman1979sample} assumes a bivariate Normal distribution for the error terms:
$$
\begin{pmatrix}
\varepsilon_i\\
\eta_i
\end{pmatrix}
 \sim \bm{N}_2\left\{\bm{0}_2=\begin{pmatrix}
0\\
0
\end{pmatrix}, \bm{\Omega}  = \begin{pmatrix}
\sigma^2 & \rho \sigma \\ \rho\sigma & 1
\end{pmatrix}\right\}.
$$
In order to achieve full identifiability, we fix the second diagonal element of $\bm{\Omega}$  at 1. The sample selection problem arises, when the error terms of the sample selection equation and the outcome equation are correlated with $\rho\neq 0$.

In order to accommodate for  heavy-tailedness, \citet{marchenko2012aheckman} propose a Heckman selection-$t$ model, replacing the error terms by a bivariate $t$ distribution with an unknown number of degrees of freedom $\nu$:
$$
\begin{pmatrix}
\varepsilon_i\\
\eta_i
\end{pmatrix}
\sim \bm{t}_2\left\{ \begin{pmatrix}
0\\
0
\end{pmatrix}, \bm{\Omega}  = \begin{pmatrix}
\sigma^2 & \rho \sigma \\ \rho\sigma & 1
\end{pmatrix}, \nu\right\}.
$$
\citet{marchenko2012aheckman} propose likelihood-based inference for the selection-$t$ model, and \citet{ding2014bayesian} proposes a Bayesian procedure to simulate the posterior distributions of the parameters.

However, the Heckman selection-$t$ model assumes that the error terms for the selection and outcome equations have the same degrees of freedom, which cannot accommodate for different heavy-tailedness in $u^*$ and $y^*$. We assume that the error terms follow an NECTD:
$$
\begin{pmatrix}
\varepsilon_i\\
\eta_i
\end{pmatrix}
\sim \bm{t}_2\left\{ \begin{pmatrix}
0\\
0
\end{pmatrix},
\bm{\Omega}  = \begin{pmatrix}
\sigma^2 & \rho \sigma \\ \rho\sigma & 1
\end{pmatrix}, 
\bm{p} =
\begin{pmatrix}
1\\
1
\end{pmatrix},
\bm{\nu} = 
\begin{pmatrix}
\nu_1\\
\nu_2
\end{pmatrix} \right \},
$$
where the numbers of degrees of freedom $\nu_1$ and $\nu_2$ are unknown.
We call it the generalized selection-$t$ mode, which takes into account many cases that cannot be described by the Heckman selection-$t$ model. For example, when $y^*_i$ is Normal, and $u_i^*$ follows a $t$ distribution with small  number of degrees of freedom, the Heckman selection-$t$ model cannot describe the heavy-tailedness of $u^*_i$ without modeling $y^*_i$ as a heavy-tailed distribution.

\subsection{Inference}
To infer the parameters in the generalized selection-$t$ model, we propose a Bayesian procedure using data augmentation and parameter expansion. We represent the error terms as
$$
\begin{pmatrix}
\varepsilon_i\\
\eta_i
\end{pmatrix}
=
\begin{pmatrix}
q_{i1}^{-1/2} & 0 \\
0& q_{i2}^{-1/2}
\end{pmatrix}
\bm{\Omega}^{1/2} \bm{Z}_i,
$$
where $q_{1i} \sim \chi^2_{\nu_1}/\nu_1,q_{1i} \sim \chi^2_{\nu_2}/\nu_2 ,  \bm{Z}_i\sim \bm{N}_2(\bm{0}_2, \bm{I}_2)$, and $(q_{1i}, q_{2i}, \bm{Z}_i)$ are mutually independent.  

For Bayesian inference,  we need to specify prior distributions for all the parameters. We choose a multivariate Normal prior for the  coefficients $(\bm{\beta}, \bm{\gamma}) \sim \bm{N}_{K+L}(\bm{\mu}_0, \bm{\Sigma}_0)$,  and Gamma priors for the degrees of freedom $\nu_i \sim \text{Gamma}(\theta_0,\phi_0)$ with shape parameter $\theta_0$ and rate parameter $\phi_0$.

In the imputation step, we first impute $(y^*_i,u^*_i)$ from Normal and truncated Normal distributions, and then draw $(q_{i1},q_{i2})$ using Metropolized Independence Samplers \citep{liu2008monte}.
In the posterior step, it is straightforward to sample the parameters due to conditional conjugacy except for the covariance matrix $\bm{\Omega}$. 
The variance of the error term in the selection equation is restricted to be 1, making the posterior distribution of the covariance matrix non-standard and difficult to sample directly. We use parameter expansion to facilitate computation, and consider the unrestricted covariance 
$$\bm{\Sigma} = \text{diag}\{1,\sigma_2\} {\ } \bm{\Omega} {\ }\text{diag}\{1,\sigma_2\}. $$
 The inverse-Wishart prior $\text{Inv-Wishart}(\nu_0,\bm{I}_2)$ for the covariance matrix $\bm{\Sigma}$ is equivalent to the priors for $(\bm{\Omega}, \sigma^2_2)$  \citep{ding2014bayesian}:
\begin{eqnarray*}
&f(\bm{\Omega}) &\propto \quad (1-\rho^2)^{-3/2}\sigma_1^{-\nu_0+3}\exp{\left\{-\frac{1}{2\sigma_1^2(1-\rho^2)}\right\}}, \\
 &\sigma_2^2 \mid \bm{\Omega} & \sim \quad \{(1-\rho^2) \chi^2_{\nu_0}\}^{-1} . 
\end{eqnarray*}
We sample $(\sigma_2, \bm{\Omega})$ jointly, and then marginalize over $\sigma_2$ by discarding their samples. 
We present the computation details  in Appendix C of the on-line supplementary materials.

\subsection{Numerical Example}
\label{sel:sim}

We generate the covariates from $x_{1i} \sim N(0,2^2), x_{2i} \sim N(0,2^2)$, and $x_{1i}$ is independent of $x_{2i}$; generate the latent outcome and selection mechanism from $y^*_i=0.5+\beta_1 x_{1i}+\epsilon_i, u^*_i=2+\gamma_1 x_{1i}+\gamma_2 x_{12}+\eta_i$, with $\beta_1=1,\gamma_1=1$ and $\gamma_2=1.5$, and 
$$
\begin{pmatrix}
\varepsilon_i\\
\eta_i
\end{pmatrix}
\sim \bm{t}_2 \left\{  
\begin{pmatrix}
0\\
0
\end{pmatrix}, \bm{\Omega}=
\begin{pmatrix}
1 & 0.3 \\
0.3& 1
\end{pmatrix}, 
\bm{p}= 
\begin{pmatrix}
1\\
1
\end{pmatrix},
 \bm{\nu}=
 \begin{pmatrix}
 30\\
 5
 \end{pmatrix}
\right\} .
$$
In the generated data set, the sample size is $3000$, with about 30\% outcomes  missing. We  apply Bayesian procedures to  the Heckman selection model, the Heckman selection-$t$ model, and the generalized selection-$t$ model.
We choose the parameters for  prior distributions as follows: $\bm{\mu}_0=\bm{0}_{K+L}$, $ \bm{\Sigma}_0 = \text{diag}\{1,\ldots,1\}/100$, $\nu_0=3$.  To investigate the sensitivity of our results to different priors,  we choose three different  priors for $\bm{\nu}$.  The prior for $\bm{\nu}$ should have  wide 95\% quantile ranges, which allows for extreme heavy-tailedness, moderate heavy-tailedness, and light-tailedness. Hence, we choose the following priors: $\text{Gamma}(1,0.1)$, $\text{Gamma}(0.5,0.05)$ and $\text{Gamma}(1.5,0.15)$, whose 95\% quantile ranges are $(0.253,36.9)$, $(0.010,50.2)$ and $(0.719,31.2)$,  respectively. We  present only the results with prior $\text{Gamma}(1,0.1)$ and give the results for other two priors in Appendix F of the on-line supplementary materials. Under different priors of $\bm{\nu}$, the parameters in the outcome equation   barely change  and the parameters in the selection equation are different. However, qualitative conclusions remain the same. In all of our later examples and applications, we run the MCMC algorithms for $5 \times 10^4$ iterations, discarding the first $10^4$ draws as a burn-in period. The results from multiple chains differ very slightly, and all of them converge with Gelman--Rubin diagnostic statistics close to 1. Therefore, we present only the results from a single chain. 

\begin{figure}
  \centering
\subfigure[Data generated from the generalized selection-$t$ model, and analyzed by the Heckman selection model (solid), the  Heckman selection-$t$ model (dotted), and the generalized selection-$t$ model (dashed). ]{
    \label{fig:selectoin:sim} %% label for second subfigure
\includegraphics[width=\textwidth]{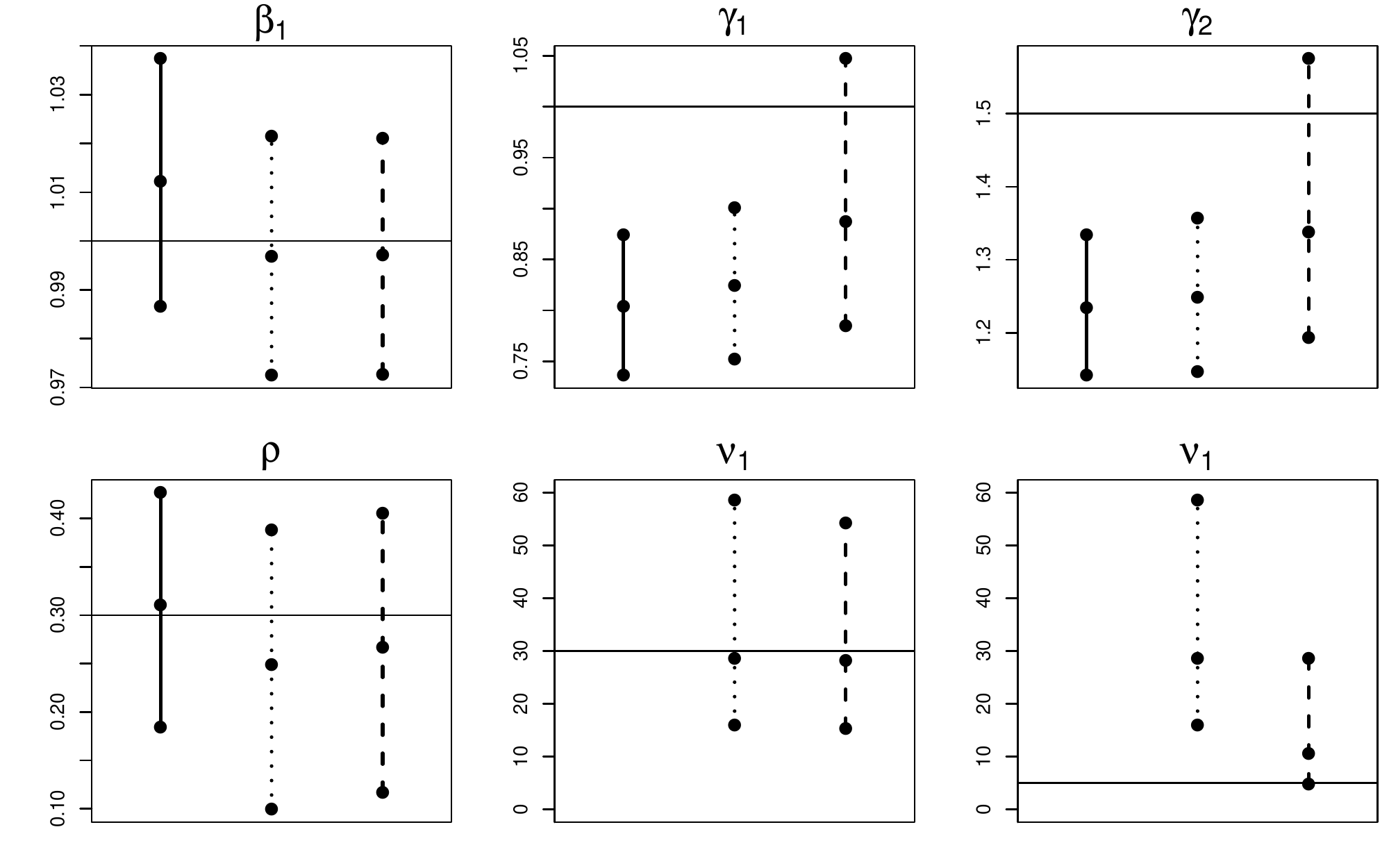}}
  \subfigure[Wage offer function analyzed by  by the Heckman selection model (solid), the  Heckman selection-$t$ model (dotted), and the generalized selection-$t$ model (dashed). A title with a subscript ``o'' or ``s'' means that the  corresponding variable is in  the outcome model or the selection model, respectively.]{
    \label{fig:selection:mroz} %% label for first subfigure
\includegraphics[width=\textwidth]{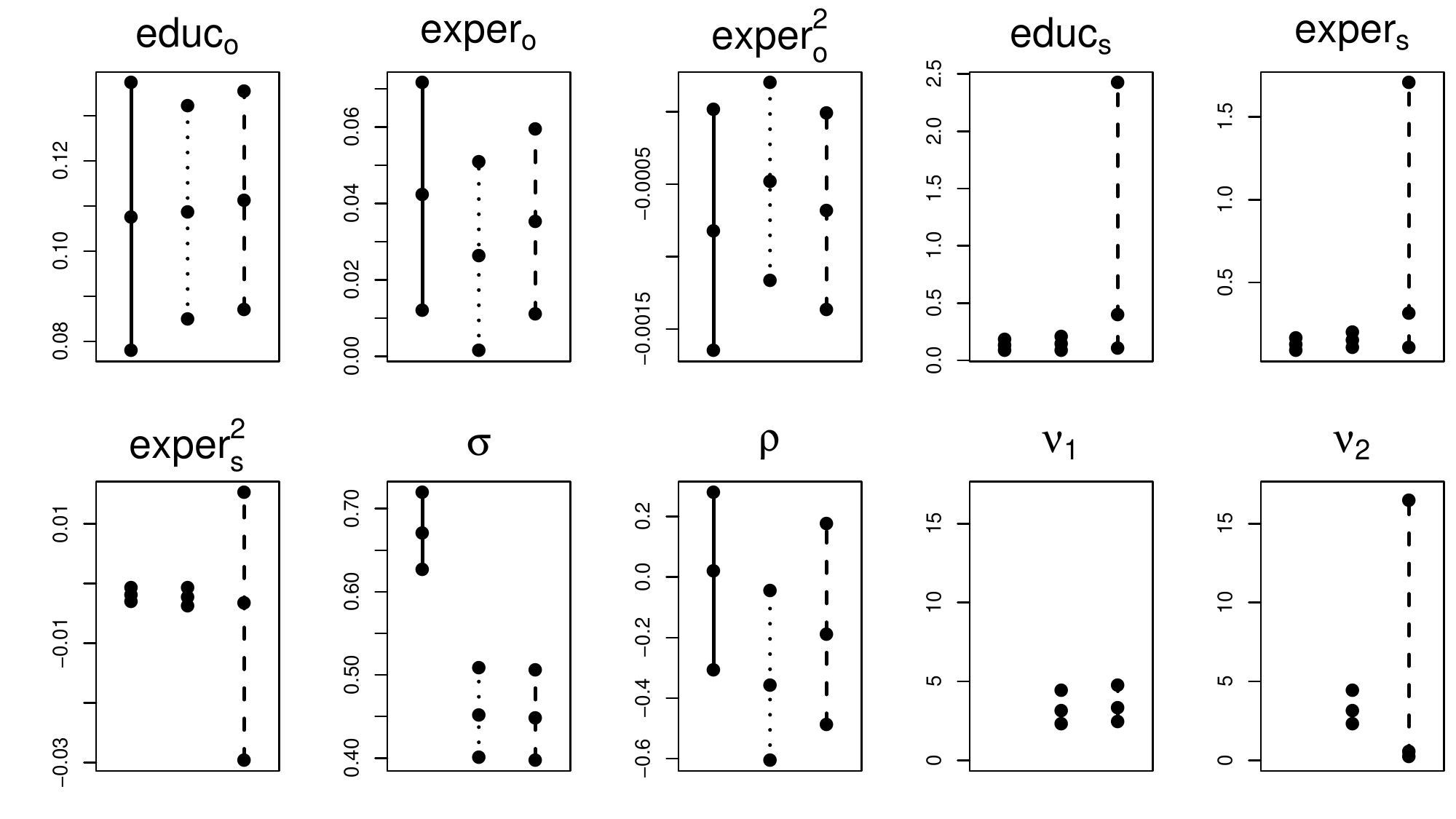}}
  \caption{Generalized selection-$t$ model.}
  \label{fig:selectionT} %% label for entire figure
\end{figure}

 Figure \ref{fig:selectoin:sim} summarizes  the posterior 2.5\%, 50\% and 97.5\% quantiles of $(\beta_1, \gamma_1, \gamma_2, \rho)$.  Under the Heckman selection model and  the Heckman selection-$t$ model, the 95\% credible intervals of $\gamma_1$ and $\gamma_2$ do not cover the true values, but under the generalized selection-$t$ model, all the 95\% credible intervals cover the true values. Thus, the simulation shows the superiority of the generalized selection-$t$ model compared with the other two models, when handling the problem of different marginal heavy-tailedness in the selection and outcome equations.

%
%\begin{table}[htp]
%\footnotesize
%\caption{Data generated from the generalized selection-$t$ model and analyzed by the Heckman selection model, the  Heckman selection-$t$ model, and the generalized selection-$t$ model. We show the Bayesian posterior medians and 95\% credible intervals (in parenthesis).}
%\begin{center}
%\begin{tabular}{rrrrrrrrr} \hline
%  True parameters           & \multicolumn{2}{c}{Heckman Selection} &    & \multicolumn{2}{c}{Heckman Selection-$t$} && \multicolumn{2}{c}{Generalized Selection-$t$} \\ \cline{1-3} \cline{5-6} \cline{8-9}
%$\beta_1=1$       & $1.012$  & $(0.987,1.037)$&  &$0.997$                           &$(0.972,1.021)$&   &    $0.997$      & $(0.973,1.021)$          \\
%$\gamma_1=1$  & $0.804$  &$(0.736,0.874)$ &  &$0.824$                           & $(0.752,0.901)$ & & $0.887$        &    $(0.784,1.047)$       \\
%$\gamma_2=1.5$&  $1.234$ & $(1.142,1.334)$&    &$1.248$                          &$(1.146,1.357) $  && $1.338$      & $(1.193,1.576)$         \\ \
%$\rho=0.3$          & $0.310$& $(0.184,0.427)$& &$0.249 $    &$(0.299,0.388)$ &    &   $0.267$      & $(0.116,0.405)$                 \\
%$\nu_1=30$     &$+\infty$& &  & $28.524$  &$(15.883,58.489)$  &   & $28.089$           & $(15.241,54.164)$         \\
%$\nu_2=5$     &$+\infty$& &  & $28.524$  &$(15.883,58.489)$ &   &$10.577$  &$(4.693,28.574)$         \\
% \hline
%\end{tabular}
%\end{center}
%\label{tab:selectT}
%\end{table}%

%\begin{figure}
%  \centering
%\includegraphics[width=\textwidth]{select_sim}
%  \caption{Data generated from the generalized selection-$t$ model and analyzed by the Heckman selection model (white), the  Heckman selection-$t$ model (light grey), and the generalized selection-$t$ model (grey). }
%  \label{fig:1} %% label for entire figure
%\end{figure}

\subsection{Application to Wage Offer Function}
 We analyze the data from \citet{mroz1987sensitivity} and \citet{wooldridge2010econometric} to estimate the wage offer function for married women. The outcome of interest is the log of wage, which are missing for 325 individuals and observed for 428 individuals. The covariates in the outcome equation are $\bm{x}=(1,\text{educ},\text{exper},\text{exper}^2)$, including education status, experience and its squared term.  The covariates in the selection equation  are $\bm{w}=(\bm{x}, \text{nwifeinc}, \text{age}, \text{kids5}, \text{kids618})$, including  income, age, number of young children and number of older children as additional covariates. Figure \ref{fig:selection:mroz} shows  the results for the Heckman selection model, the Heckman selection-$t$ model, and the generalized selection-$t$ model. We do not show the results for the covariates other than $\bm{x}$, because our focus is on $\bm{x}$ and the results of other covariates have the same pattern as the results of $\bm{x}$.

Under the generalized selection-$t$ model, the posterior distributions of the coefficients in the selection model are more dispersed. This is because in the generalized selection-$t$ model, the heavy-tailedness of  the selection model does not depend on the outcome model, and thus the information for the selection model from the data is less than the other two models. The qualitative conclusion about these coefficients remains the same in all the three models.
The posterior median of the number of degrees of freedom of the  selection equation is 0.544 under the generalized selection-$t$ model, implying severe heavy-tailedness in the sample selection process. In addition, the  numbers of degrees of freedom for the selection and outcome equations differ dramatically. The credible intervals of these two numbers of degrees of freedom have some overlap because of large variability of $\nu_2$. However, the 95\% credible interval of $\nu_1$ covers only $6.6\%$ of credible interval of $\nu_2$, which suggests great difference between these two degrees of freedom. Under the Heckman selection-$t$ model, the 95\% credible interval  of $\rho$  does not cover zero, which indicates the existence of sample selection. However, under the generalized selection-$t$ model, the posterior distribution of $\rho$ covers zero, showing weak evidence for the sample selection effect. The different  conclusions about the sample selection might be attributed to the different numbers of degrees of freedom in the selection and outcome equations.   Moreover, under the Heckman selection model, there is no evidence of the sample selection effect either. Thus the sample selection effect might be induced by the restriction on the number of degrees of freedom in the MTD as illustrated in Example 1.  For the coefficients of the outcome equation, the three models generate  similar results, but 
for the coefficients of the selection equation, the three models differ in the scale of the estimates. These differences might also be due to the different posterior distributions of the numbers of degrees of freedom of the selection equation.

\section{Generalized Multivariate Robit Model}

\subsection{Model}
Logistic and Probit models are widely used to model binary data in practice. However, analyses based on Logistic and Probit models are not robust to outliers, because they can be represented as latent linear models with Logistic and Normal error terms.
Robit models, with $t$ distributed error terms in the latent linear models \citep{liu2004robit,albert1993bayesian,mudholkar1978remark}, allow for flexible modeling of data with heavy tails.

We first introduce the multivariate Robit model with unknown number of degrees of freedom. 
The observed variables $\bm{y}_i=(y_{i1},\ldots, y_{ip})^\top$ are truncated versions of latent variables $\bm{y}^*_i=(y^*_{i1},\ldots, y^*_{ip})^\top$ via $y_{ij}=I(y^*_{ij}>0)$, with the latent variables modeled as 
\begin{eqnarray*}
\bm{y}^*_i = \bm{x}_i \bm{\beta}+\bm{\varepsilon_i}, 
\end{eqnarray*}
 where $ \bm{\varepsilon_i} \sim \bm{t}_p(\bm{0}_p, \bm{\Omega}, \nu),$ and $\bm{x}_i$  is a known $p\times K$  design matrix. In order to achieve full identification, we restrict the diagonal elements of $\bm{\Omega}$ to be one. Similar to the generalized selection-$t$ model, we can replace the distribution of the error terms by an NECTD with an unknown vector of numbers of degrees of freedom $\bm{\nu}=(\nu_1,\ldots,\nu_s)^\top$, i.e., $\bm{\varepsilon_i} \sim \bm{t}_p(\bm{0}_p, \bm{\Omega}, \bm{p}, \bm{\nu})$. This model can describe cases where elements of the latent variable $\bm{y}^*_i$ have different marginal heavy-tailedness.

\subsection{Inference}
To infer the parameters in the generalized Robit model, we propose a Bayesian procedure using data augmentation and parameter expansion. We represent the error terms as
$$
\bm{\varepsilon}_i \sim \bm{Q}_i^{-1/2}   \bm{\Omega}^{1/2}  \bm{Z}_i, \quad \bm{Z}_i\sim \bm{N}_p(\bm{0}, \bm{I}_p), \quad \{ q_{ij}\sim \chi^2_{\nu_j}/\nu_j: j=1, \ldots, s\},
$$
where $\bm{Q}_i  = \text{diag}  \{q_{i1} \bm{I}_{p_1}, \ldots, q_{is} \bm{I}_{p_s}\}$ is a block diagonal matrix with $\sum_{j=1}^{s}p_j=p$, and the $q_{ij}$'s and $\bm{Z}_i$'s are mutually independent.
%. Note that $\{q_{ij}: j=1, \ldots, s\}$ are mutually independent, and  $q_{ij} \ind \bm{Z}_i$ for $j=1, \ldots, s$.

For Bayesian inference,  we need to specify prior distributions for all the parameters. We choose a multivariate Normal prior for the coefficients $\bm{\beta} \sim \bm{N}_{K}(\bm{\mu}_0, \bm{\Sigma}_0)$, and Gamma priors for the degrees of freedom $\nu_i \sim \text{Gamma}(\theta_0,\phi_0)$.

In the imputation step, we treat $\bm{y}^*_{i}$'s and $\bm{Q}_i$'s as missing data. Except for $\bm{\Omega}$, the posterior distributions of the parameters have conditional conjugate forms. The diagonal elements of $\bm{\Omega}$ are restricted to be 1 for identification, making the posterior distribution of the covariance matrix non-standard and difficult to sample directly. We solve this problem by using parameter expansion, and consider the unrestricted covariance 
$$
\bm{\Sigma} = \text{diag}\{  d_1,\ldots,d_p \}~ \bm{\Omega} ~ \text{diag}\{ d_1,\ldots,d_p\} . 
$$ 
The inverse-Wishart prior $\text{Inv-Wishart}(\nu_0,\bm{I}_p)$ for the covariance matrix $\bm{\Sigma}$ is equivalent to the priors for $(\bm{\Omega}, d_1,\ldots, d_p)$:
\begin{eqnarray*}
f(\bm{\Omega}) &\propto& | \bm{\Omega} |^{-(\nu_0+p+1)/2} \left (\prod_i \omega^{ii}\right)^{-\nu_0/2}, \\
 d_i^2 \mid \bm{\Omega} & \sim&  \omega^{ii}/\chi^2_{\nu_0},
\end{eqnarray*}
where $\omega^{jj}$ is the $(j,j)$-th element of $\bm{\Omega}^{-1}$ \citep{barnard2000modeling}. 
 We sample $(d_1,\ldots,d_p,  \bm{\Omega})$ jointly, and then marginalize over the $d_i$'s by discarding their samples. We present the computation details in Appendix D of the on-line supplementary materials.

\subsection{Numerical Example}

We generate the covariates from $x_{i1} \sim N(0,1), x_{i2} \sim N(0,1)$; generate the latent outcome 
$\bm{y}^*_i= (y^*_{i1}, y^*_{i2} )^\top$ from $y^*_{i1}=\beta_0+\beta_1 x_{i1}+\varepsilon_{i1}, y^*_{i2}=\beta_0+\beta_1 x_{i2}+\varepsilon_{i2}$, with $\beta_0=0.5, \beta_1=1$, and 
$$
\begin{pmatrix}
\varepsilon_{i1}\\
\varepsilon_{i2}
\end{pmatrix}
\sim \bm{t}_2 \left\{  
\begin{pmatrix}
0\\
0
\end{pmatrix}, \bm{\Omega}=
\begin{pmatrix}
1 & 0.2 \\
0.2& 1
\end{pmatrix}, 
\bm{p}=
\begin{pmatrix}
1\\
1
\end{pmatrix},
\bm{\nu}=
\begin{pmatrix}
5\\
30
\end{pmatrix}
\right \}.
$$
The observed outcomes are $y_{i1}=I(y^*_{i1}>0)$ and $y_{i2}=I(y^*_{i2}>0)$.  The sample size is 3000 in our generated data set. 
We choose the parameters for  prior distributions as follows: $\bm{\mu}_0=\bm{0}_{K}, \bm{\Sigma}_0 = \text{diag}\{1,\ldots,1\}/100, \nu_0=p+1, \theta_0=1$, and $\phi_0=0.1$. 
In Appendix F of the on-line supplementary materials, we conduct sensitivity analysis  and show that the results are not sensitive to different priors of $\nu$. 

\begin{figure}
  \centering
\subfigure[Data generated from the generalized Robit model, and analyzed by the Probit model (solid), the Robit model (dotted), and the generalized Robit model (dashed). ]{
    \label{fig:robit:sim} %% label for second subfigure
\includegraphics[width=\textwidth]{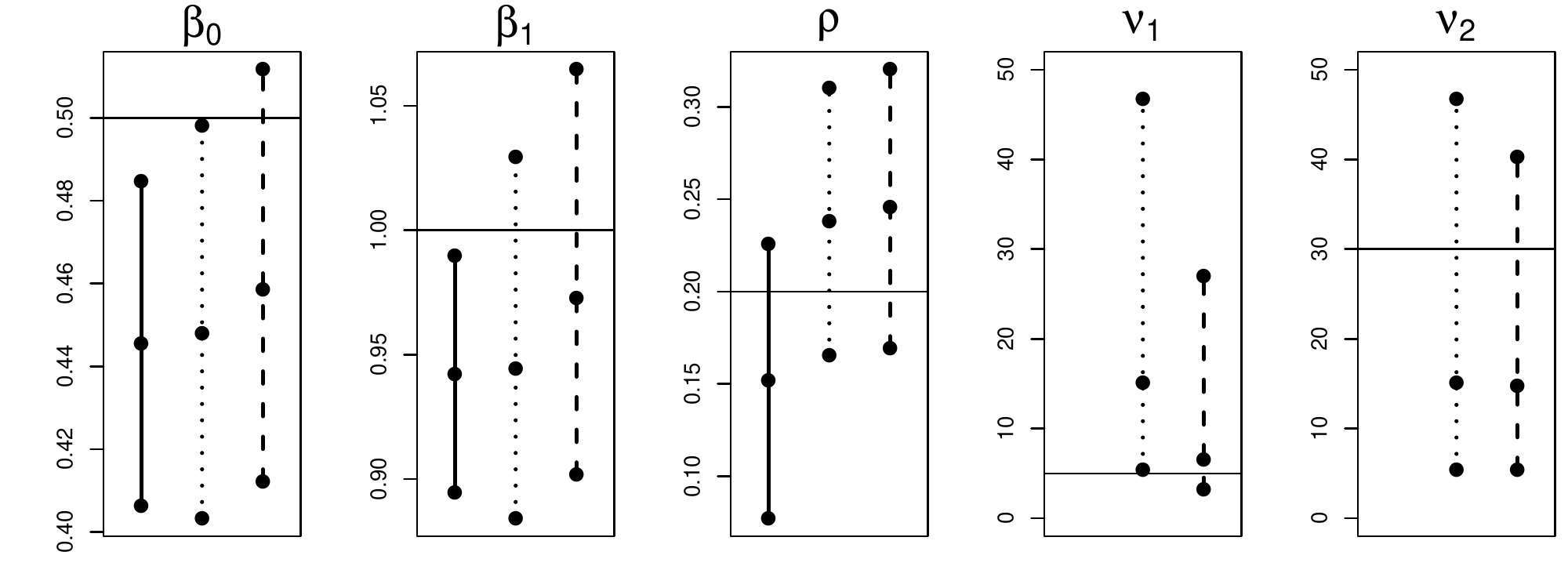}}
  \subfigure[The flu shot experiment data analyzed by the Probit model (solid), the Robit model (dotted)  and the generalized Robit model (dashed).]{
    \label{fig:robit:hirano} %% label for first subfigure
\includegraphics[width=\textwidth]{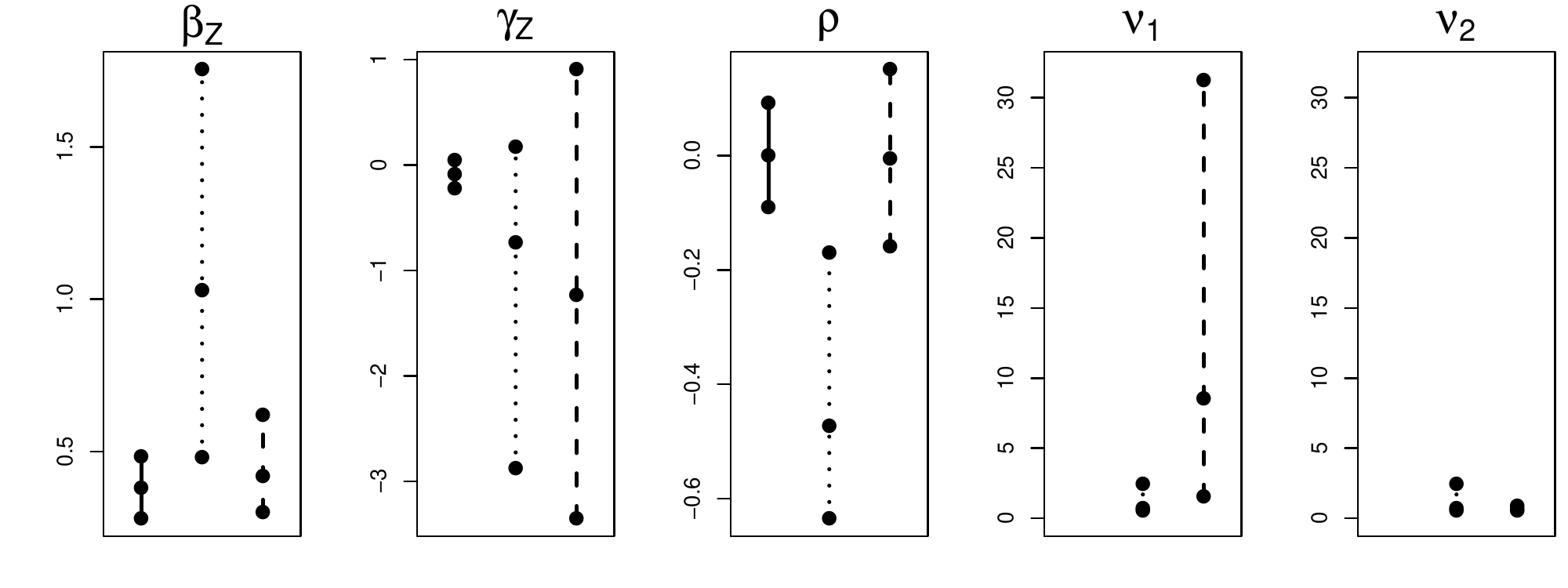}}
  \caption{Generalized multivariate Robit model.}
  \label{fig:robit} %% label for entire figure
\end{figure}

We  apply the Bayesian procedures for the Probit model, the  Robit model, and the generalized Robit model. The boxplots in Figure \ref{fig:robit:sim} summarize the posterior quantiles of $(\beta_0, \beta_1, \rho,\nu_1,\nu_2)$. 
Under the Probit and the Robit models, the 95\% credible intervals of $\beta_0$ do not cover the true value; under the generalized Robit model, the 95\% credible interval of $\beta_0$ covers the true value.

%Under the Probit and the Robit models, the posterior distributions of $\beta_0$ are quite away from the true value (in fact, the 95\% credible intervals do not contain 0); under the generalized Robit model, the  posterior distributions of $\beta_0$ cover the true value.  
 
%\begin{table}[htp]
%\footnotesize
%\caption{Data generated from the generalized Robit model and analyzed by the Probit model, the Robit model, and the generalized Robit model. We show the results for Bayesian posterior medians and 95\% credible intervals (in parenthesis).}
%\begin{center}
%\begin{tabular}{rrrrrrrrr} \hline
% True parameters        & \multicolumn{2}{c}{Probit}&     & \multicolumn{2}{c}{Robit} & &\multicolumn{2}{c}{Generalized Robit} \\ \cline{1-1} \cline{2-3} \cline{5-6} \cline{8-9}
%
%
%$\beta_0=0.5$    &$0.445$   &$(0.406,0.485)$ &  &$0.448$                           & $(0.403,0.498)$&  & $0.459$        &    $(0.412,0.512)$       \\
%$\beta_1=1$ &  $0.942$  & $(0.894,0.990)$ &  &$0.944$                          &$(0.884,1.029)$ & & $0.972$      & $(0.902,1.064)$         \\
%$\rho=0.2$         &   $0.152$&  $(0.077,0.226)$& &$0.238$                           &$(0.165,0.310)$   &    &$0.246$      & $(0.169,0.320)$          \\
%$\nu_1=5$     &$+\infty$&&   & $14.085$  &$(5.468,44.592)$&     & $6.490$           & $(3.162,26.978)$         \\
%$\nu_2=30$     &$+\infty$&&   & $14.085$  &$(5.468,44.592)$&   &$14.714$  &$(5.393,40.282)$         \\
% \hline
%\end{tabular}
%\end{center}
%\label{tab:robit}
%\end{table}%

%\begin{figure}
%  \centering
%\includegraphics[width=\textwidth]{robit_sim}
%  \caption{Data generated from the generalized Robit model and analyzed by the Probit model (white), the Robit model (light grey), and the generalized Robit model (grey). }
%  \label{fig:robit:sim} %% label for entire figure
%\end{figure}

\subsection{Application to a Flu Shot Experiment }
We reanalyze the data in  \citet{hirano2000assessing}.
In this study, physicians were randomly selected to receive a letter encouraging them to inoculate patients at risk for flu. The treatment of interest is the actual flu shot, and the outcome is an indicator for flu-related hospital visits. However, some patients did not comply with their assignments. Let $Z_i$ be the indicator of encouragement to receive flu shot, with $Z_i=1$ if patient $i$'s physician received the encouragement letter, and  $Z_i=0$ otherwise. Let $D_i$ be the treatment received, with $D_i=1$ if patient $i$ received the flu shot, and $D_i=0$ otherwise.
Let $Y_i$ be the outcome, with $Y_i=1$ if patient $i$ subsequently experienced a flu-related hospitalization during the winter, and $Y_i=0$ otherwise.  Let $X_i$ be the pretreatment covariates. We assume the following generalized Robit model for the joint value of $(D,Y)$:
\begin{eqnarray*}
&&Y^*_i=\beta_0+\beta_{Z}Z_i+\beta_X X_i +\varepsilon_{1i},  \quad Y_i=I(Y^*_i>0),\\
&&D^*_i=\gamma_0+\gamma_{Z}Z_i+\gamma_X X_i+\varepsilon_{2i}, \quad  D_i=I(D^*_i>0),\\
&&\begin{pmatrix}
\varepsilon_{1i}\\
\varepsilon_{2i}
\end{pmatrix}
\sim \bm{t}_2 \left\{  
\begin{pmatrix}
0\\
0
\end{pmatrix}, \bm{\Omega}=
\begin{pmatrix}
1 & \rho \\
\rho& 1
\end{pmatrix}, 
\bm{p}= 
\begin{pmatrix}
1\\
1
\end{pmatrix},
\bm{\nu}=
\begin{pmatrix}
\nu_1\\
\nu_2
\end{pmatrix}
\right\} .
\end{eqnarray*}
Figure \ref{fig:robit:hirano} shows the results for the Probit model,  the Robit model,  and the generalized Robit model. Under the Robit model, the posterior median of the number of degrees of freedom is $0.676$, which has strong evidence of heavy-tailedness. However, the posterior distributions of the two numbers of degrees of freedom in the generalized Robit model differ greatly, which makes the result very different from that of the Robit model. Ignoring this difference might lead to biased inference.

In the causal inference literature, the randomly assigned $Z$ in the encouragement design is often used as an instrumental variable for identifying causal effect of the treatment received $D$ on the outcome $Y$ \citep{hirano2000assessing,angrist1996identification}. The instrumental variable $Z$ must first satisfy the condition that $Z$ and $D$ are correlated. However, the 95\% credible interval of $\gamma_Z$  is covers 0, indicating that the correlation between $Z$ and $D$ is weak. Thus, $Z$ is a very weak instrument. The instrumental variable $Z$ must also satisfy the exclusion restriction assumption, i.e., $Z$ affects $Y$ only through $D$.
Under all the three models, however, the 95\% credible interval of $\beta_Z$ does not cover zero, which means that the intention-to-treat effect of the encouragement on the outcome is positive. Combining this with the fact that $\gamma_Z$ is near zero, we suspect that the encouragement has a ``direct effect'' on the outcome not through $D$, and thus the exclusion restriction assumption does not hold.

Under the generalized selection-$t$ model, the posterior median of $\rho$ is very close to zero, and therefore it is plausible to assume that  $D$ and $Y$ are independent conditional on $	Z$ and $X$.  The estimate of $\rho$ is similar under the Probit model but is different under the Robit model. This might be induced by the restriction of the degrees of freedom in the MTD.

%\begin{table}[htp]
%\footnotesize
%\caption{The flu shot experiment. We show the Bayesian posterior medians and $95\%$ confidence intervals (in parenthesis) of the Probit model, the Robit model  and the generalized Robit model.}
%\begin{center}
%\begin{tabular}{rrrrrrrrr} \hline
%             & \multicolumn{2}{c}{Probit} &    & \multicolumn{2}{c}{Robit}& & \multicolumn{2}{c}{Generalized Robit} \\ \cline{1-3}  \cline{5-6} \cline{8-9} 
%
%
%$\beta_0$    &$-1.636$   &$(-1.957,-1.321)$ &  &$-3.431$                           & $(-4.914,-1.928)$&  & $-1.773$        &    $(-2.438,-1.392)$       \\
%$\beta_Z$ &  $0.381$ &   $(0.281,0.483)$ &&$1.028$                          &$(0.479,1.754)$&  & $0.420$      & $(0.302,0.619)$         \\
%$\gamma_0$        &$-1.264$& $(-1.650,-0.891)$& &$-5.658$    &$(-11.692,-1.460)$&     &   $-8.619$      & $(-14.745,-3.421)$                 \\
%$\gamma_Z$     &$-0.089$& $(-0.224,0.044)$&  &$-4.742$     &$(-2.880,0.167)$&    &$-1.231 $        &  $(-3.349,0.904)$                     \\ \hline
%$\rho$         &  $-0.001$  &$(-0.091,0.091)$&   &$-0.474$                           &$(-0.634,-0.170)$   &   & $-0.005$      & $(-0.159,0.150)$          \\
%$\nu_1$     &$+\infty$& &  & $0.676$  &$(0.535,2.448)$   &  & $8.558$           & $(1.521,31.224)$         \\
%$\nu_2$     &$+\infty$&  & & $0.676$  &$(0.535,2.448)$&   &$0.664$  &$(0.536,0.849)$         \\
% \hline
%\end{tabular}
%\end{center}
%\label{tab:apl:robit}
%\end{table}%

\section{Generalized  Linear $t$ Mixed-Effects Model}
\subsection{Model}
Linear mixed-effects models \citep{hartley1967maximum} are popular for analyzing repeated measurements, which arise in many areas such as agriculture, biology, economics, and geophysics.  For a continuous response, \citet{laird1982random} propose the following linear mixed-effects model:
\begin{eqnarray*}
\bm{y}_i=\bm{x}_i \bm{\beta}+\bm{z}_i \bm{b}_i+\bm{\epsilon}_i,
\end{eqnarray*}
where  $\bm{y}_i=(y_{i1},\ldots,y_{in_i})^\top $ is the outcome vector; $\bm{x}_i$ and $\bm{z}_i$ are known $n_i\times K$ and $n_i\times L$ design matrices corresponding to the $K$-dimensional fixed effects vector $\bm{\beta}$ and the $L$-dimensional random effects vector $\bm{b}_i$, respectively; $\bm{\varepsilon}_i$ is an $n_i$-dimensional vector of within-subject errors independent of $\bm{b}_i$. The $\bm{b}_i$'s are  independent
 with distribution $\bm{N}_L(0,\bm{\Omega})$, and $\bm{\varepsilon}_i$'s are  independent  with distribution $\bm{N}_{n_i}(0,\bm{\Lambda}_i)$. Thus, the random effects and the within-subject errors follow a multivariate Normal distribution:
$$
\begin{pmatrix}
\bm{b}_i\\
\bm{\varepsilon}_i
\end{pmatrix}
\sim \bm{N}_{L+n_i}\left\{
\begin{pmatrix}
0\\
0
\end{pmatrix} ,
\begin{pmatrix}
 \bm{\Omega} & 0 \\
0& \bm{\Lambda}_i
\end{pmatrix}\right\}.
$$
Here,  the $L \times L$ matrix $\bm{\Omega}$ and $n_i \times n_i$ matrix $\bm{\Lambda}_i$ are non-singular covariance matrices. The matrix $\bm{\Omega}$ may be unstructured or structured, but $\bm{\Lambda}_i$ is generally parametrized in terms of a small number of parameters that do not change with $i$.
  \citet{pinheiro2001efficient} replace the multivariate Normal distribution by a MTD with an unknown degrees of freedom $\nu$:
 $$
\begin{pmatrix}
\bm{b}_i\\
\bm{\varepsilon}_i
\end{pmatrix}
\sim \bm{t}_{L+n_i}\left\{
\begin{pmatrix}
0\\
0
\end{pmatrix} ,
\begin{pmatrix}
 \bm{\Omega} & 0 \\
0& \bm{\Lambda}_i
\end{pmatrix}, \nu\right\}.
$$

Thus, they assume that the marginal distributions of random effects and within-subject errors have the same number of degrees of freedom. To allow for the different heavy-tailedness for random effects and within-subject errors, we replace the MTD by an NECTD:
 $$
\begin{pmatrix}
\bm{b}_i\\
\bm{\varepsilon}_i
\end{pmatrix}
\sim \bm{t}_{L+n_i}\left\{
\begin{pmatrix}
0\\
0
\end{pmatrix} ,
\begin{pmatrix}
 \bm{\Omega} & 0 \\
0& \bm{\Lambda}_i
\end{pmatrix}, 
\bm{p} =
\begin{pmatrix}
L\\
n_i
\end{pmatrix},
 \bm{\nu}=
 \begin{pmatrix}
 \nu_1\\
 \nu_2
 \end{pmatrix}
 \right\},
$$
where $\nu_1$ and $\nu_2$ are the numbers of degrees of freedom for random effects and within-subject errors, respectively.

\subsection{Inference}

We propose a Bayesian procedure to infer the parameters in the generalized linear $t$ mixed-effects model. For simplicity, we assume $\bm{\Lambda}_i$ to be diagonal, i.e., $\bm{\Lambda}_i=\sigma^2 \bm{I}_{n_i}$.
We represent the random effects and the within-subject errors as
\begin{eqnarray*}
\bm{b}_i\mid q_{i1} \sim \bm{N}_L(\bm{0}, \bm{\Omega}/q_{i1}), \quad q_{i1} \sim \chi^2_{\nu_1}/\nu_1,\\ 
 \bm{\epsilon}_i\mid q_{i2} \sim \bm{N}_{n_i}(\bm{0}, \sigma^2 \bm{I}_{n_i}/q_{i2}) , \quad q_{i2} \sim \chi^2_{\nu_2}/\nu_2.
\end{eqnarray*}
For Bayesian inference,  we need to specify prior distributions for all the parameters. We choose a multivariate Normal prior for the coefficients $(\bm{\beta}, \bm{\gamma}) \sim \bm{N}_{K}(\bm{\mu}_0, \bm{\Sigma}_0)$,  Gamma priors for the numbers of degrees of freedom $\nu_i \sim \text{Gamma}(\theta_0,\phi_0)$, and an inverse-Wishart prior for the covariance matrix of the random effects $\bm{\Omega} \sim \text{Inv-Wishart}(\nu_0, \bm{I}_L)$. To guarantee a proper posterior distribution, we choose  $\sigma^2 \sim  \text{Inv-Gamma}(0.5,0.1)$ as the prior for the variance of the within-subject errors. Under these prior distribution choices, all
the conditional distributions of the latent variables and model parameters are standard and straightforward to sample.
 We present the computation details in Appendix E of the on-line supplementary materials.

\subsection{Numerical Example}

In our simulation, we choose $n_i=2, \bm{\Lambda}_i= \bm{I}_2, \bm{\beta}=(0.5,1,-0.5)^\top $, and  
$
\bm{\Omega}=
\begin{pmatrix}
1 & 0.5 \\
0.5& 1
\end{pmatrix};
$
and generate all the elements of $\bm{x}_i$ and $\bm{z}_i$ from standard Normal distributions.  We choose the parameters for  prior distributions as follows: $\bm{\mu}_0=\bm{0}_{K}, \bm{\Omega}_0 = \text{diag}\{1,\ldots,1\}/100, \nu_0=L+1, \theta_0=1$, and $\phi_0=0.1$. In Appendix F of  the on-line supplementary materials, we conduct sensitivity analysis  and show that the results are not sensitive to different priors of $\bm{\nu}$.

\begin{figure}
  \centering
\subfigure[Data generated from the generalized $t$ linear mixed-effects model, and analyzed by the linear mixed-effects model (solid),  the linear $t$ mixed model model (dotted) and the generalized  linear $t$ mixed-effects model (dashed).  ]{
    \label{fig:lmm:sim} %% label for second subfigure
\includegraphics[width=\textwidth]{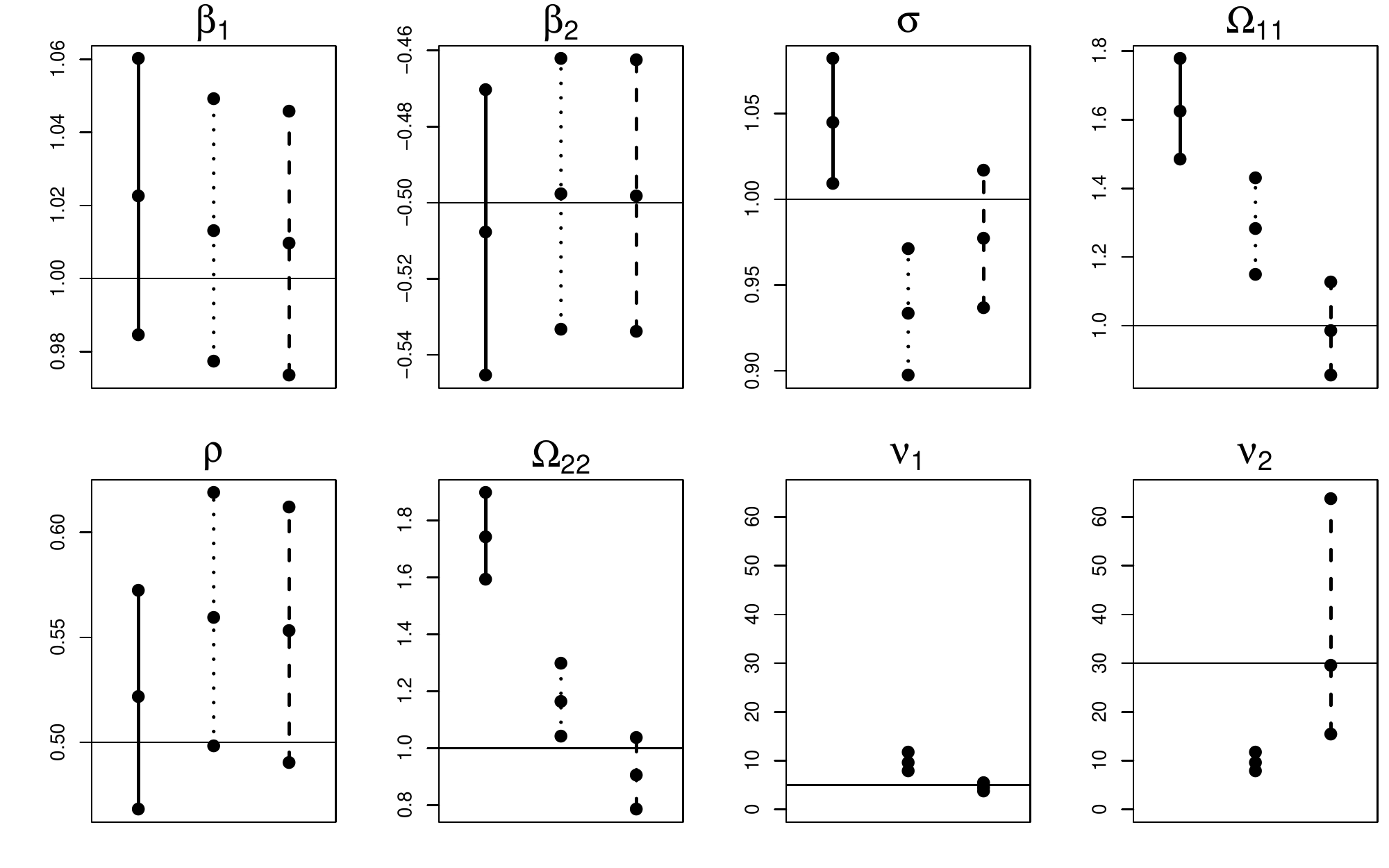}}
  \subfigure[The Framingham study example analyzed by  the linear mixed-effects model (solid),  the linear $t$ mixed model model (dotted) and the generalized  linear $t$ mixed-effects model (dashed).]{
    \label{fig:lmm:framingham} %% label for first subfigure
\includegraphics[width=\textwidth]{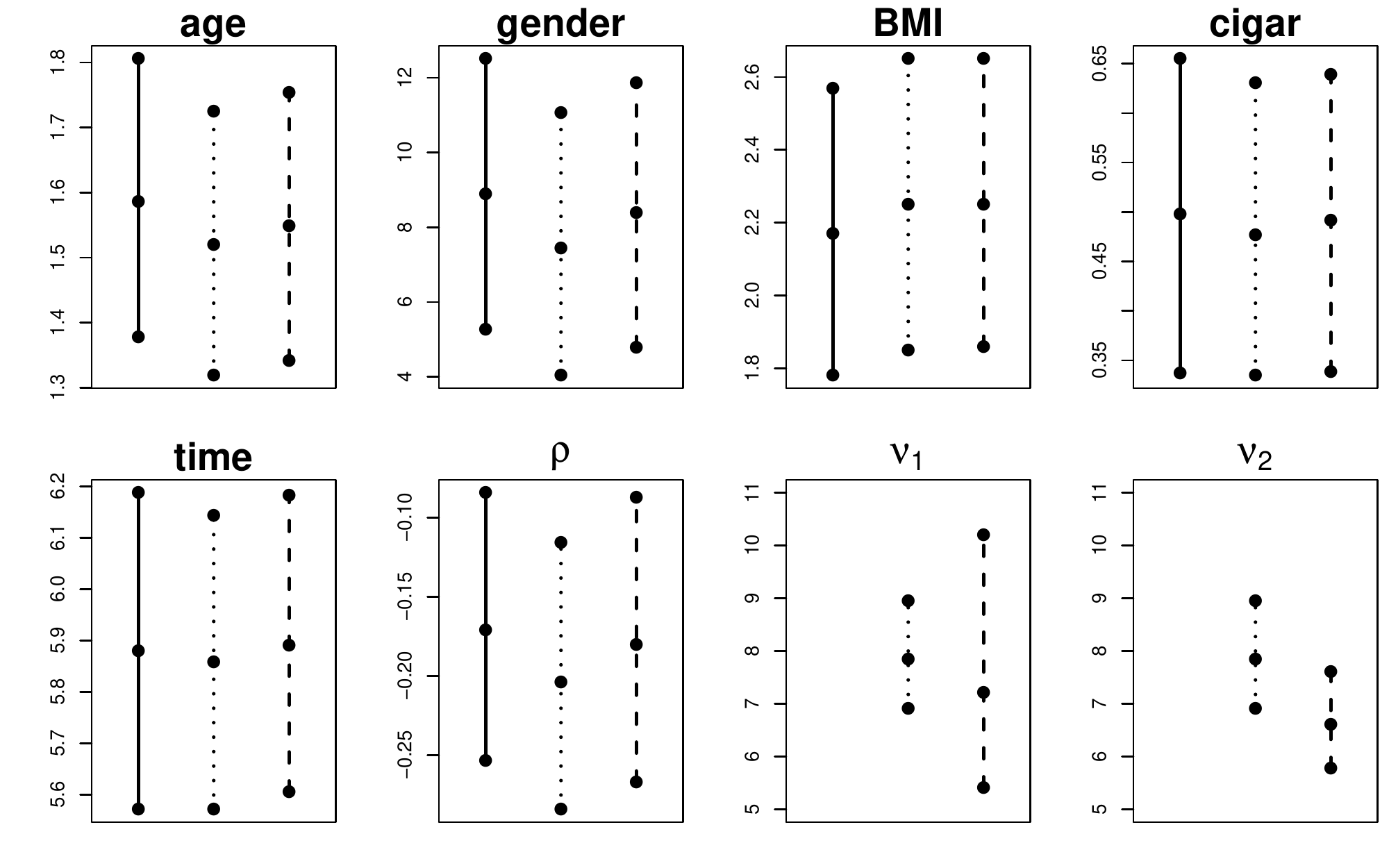}}
  \caption{Generalized linear $t$ mixed-effects model.}
  \label{fig:robit} %% label for entire figure
\end{figure}

 We apply the Bayesian procedures to the linear mixed-effects model, the linear $t$ mixed-effects model, and the generalized linear $t$ mixed-effects model. The boxplots in Figure \ref{fig:lmm:sim} summarizes the posterior distributions of the parameters. Under these three models, the posterior distributions of the coefficients  are very close. This happens because a $t$ distribution  may well approximate the linear combination of two $t$ distributions (with different numbers of degrees of freedom). Therefore, the heterogeneity of the numbers of degrees of freedom does not change the estimates of  the regression coefficients too much. However, under the linear $t$ mixed-effects model, the $95\%$ credible intervals of both the covariance matrix  of the random effects and the variance of the  within-subject errors do not contain the true values.

\subsection{Application to the Framingham Study}

We analyze the data from the Framingham study \citep{dawber1951epidemiological}, which is a long term follow-up study to identify the relationship between various risk factors and diseases. The data on various aspects have been and continue to be collected every two years on a cohort of individuals. The outcomes are the serum cholesterol levels at the baseline and then every two years through year 10. The covariates include the age of the individual when they entered the study,  gender, body mass index (BMI) at the baseline, and the number of cigarettes the individual smoked per day at the baseline. 

We present the results for  the linear mixed-effects model, the linear $t$ mixed-effects model, and the  generalized linear $t$ mixed-effects model  in Figure \ref{fig:lmm:framingham}. 
The posterior  medians of the two numbers of degrees of freedom are close, which provides weak evidence for different $\nu$'s, and thus the two models give very similar estimates of the parameters.
The 95\% intervals of all the covariates do not contain zero, which indicates that the serum cholesterol level is positively related to the age, gender, BMI, and the number of cigarettes  the individual smoked per day at the baseline. Because the 95\% credible interval of time does not contain zero, we conclude that the serum cholesterol level increases over time.

\section{Discussion}
\label{sec:conc}
%This paper generalizes the traditional multivariate $t$ distributions to a class of non-elliptically contoured multivariate $t$ distributions, which can accommodate for different marginal heavy-tailedness.  We also generalize the Heckman selection model, the Robit model, and the linear mixed-effects model using the non-elliptically contoured multivariate $t$ distributions.  %We  infer the parameters of these models efficiently  by using Bayesian procedures based on data augmentation and parameter expansion. 
%In the applications of the Heckman selection model and the Robit model, we find  evidence of heterogeneity  of heavy-tailedness. The conclusions differ under different distributional assumptions of the error terms. In the application of the linear mixed-effects model, there is no substantiated evidence of different heavy-tailedness, and the conclusions are very similar under both the generalized linear $t$ mixed-effects model and the linear $t$ mixed-effects model.

In previous sections, we assume that $s$ and $\bm{p}$ are known. This is reasonable in  the general Heckman selection-$t$, Robit, and linear mixed-effects models. For example, the reason for using the new $t$ distribution in the generalized selection model is to accommodate different tail behaviors of the selection and outcome equations, in which case $s=2$ and $\bm{p}=(1,1)$. However, there may be other scenarios in which  $s$ and $\bm{p}$ are unknown. To deal with this,  \citet{finegold2014robust} proposed the Dirichlet $t$-distribution for graphical models.  It is an interesting topic to extend it to general models.

We choose Gamma priors for the numbers of degrees of freedom and conduct sensitivity analysis with different hyperparameters. Alternatively, \citet{roy2014efficient} and \citet{roy2015efficient} suggested empirical Bayes methods, and others suggested using discrete priors on $\nu$ \citep{liu2004robit, koenker2009parametric}.
%For the Robit link, \citet{roy2014efficient} showed that it is important to efficiently estimate $\nu$ and considered an empirical Bayes methods for estimating it, which is also suggested in the analysis of spatial data \citep{roy2015efficient}. We can also consider discrete priors on $\nu$ with different levels of $\nu$ accounting for different heavy-tailedness. For more discussion, please see \citep{liu2004robit,roy2014efficient, koenker2009parametric}. 
In practice, researchers may also need to investigate the sensitivity of their results to different prior distributions on other parameters before making scientific conclusions.

For models based on $t$ errors, the basic data augmentation algorithm may suffer from slow convergence \citep{roy2010monte,roy2012convergence}. Often parameter expansion data augmentation algorithm may improve the performance of the data augmentation algorithm without much extra computational burden \citep[e.g.,][]{van2001art, ding2014bayesian}. Therefore, it is also interesting to develop more efficient data augmentation algorithm for our proposed models.

\section*{References}

\bibliography{Bibliography-GMVT}

%
%In practice, we can focus only on some specific values of $\nu$'s and  assume $\nu$'s to take values in a finite subset, e.g., $(0.5,1 , 2, 7, 30, 50)$. When $\nu=0.5$, the distribution is severely heavy-tailed; when $\nu=7$, the distribution is very similar to the standard Logistic distribution \citep{liu2004robit}; when $\nu=50$, the distribution barely differs from a Normal distribution. Thus, the discrete prior focuses on some distributions with significantly different heavy-tailedness, and makes the MCMC algorithm converge more quickly.

\section*{Appendix A: Properties of the NECTD}
We present some properties of the NECTD and take the bivariate NECTD as an example for derivation.
 
We first give the moments of the NECTD. Suppose $p=2$, $p_1=p_2=1$, $\bm{\mu}=(0,0)^\top$ and $\bm{\nu}=(\nu_1,\nu_2)^\top$. For simplicity, we discuss the standard form with
$$
\bm{\Sigma}=\begin{pmatrix}
1 & \sin(\theta) \\
\sin(\theta)& 1
\end{pmatrix}.
$$
Let $Z_1$ and $Z_2$ denote two independent standard Normal random variables.
We can write $X_1$ and $X_2$ in terms of $Z_1$ and $Z_2$:
\begin{eqnarray*}
X_1=\sqrt{\frac{\nu_1}{q_1}}Z_1, \quad X_2=\sqrt{\frac{\nu_2}{q_2}}\{Z_1 \sin(\theta)+Z_2 \cos(\theta)\}.
\end{eqnarray*}
Denote
\begin{eqnarray*}
C(n,k)=\frac{n!}{k!(n-k)!}, \quad
f(n)= \begin{cases}
   0 &\mbox{if $n$ is odd},\\
   (n-1)!! &\mbox{if $n$ is even}.
   \end{cases}
\end{eqnarray*}
Using Newton's binomial theorem,
we  have
\begin{eqnarray*}
&&E(X_1^{r_1}X_2^{r_2}\mid q_1,q_2)=\nu_1^{-r_1/2}\nu_2^{-r_2/2}q_1^{-r_1/2}q_2^{-r_2/2}E\left \{Z_1^{r_1}(Z_1\sin(\theta)+Z_2\cos(\theta))^{r_2}\right \}\\
&=& \nu_1^{-r_1/2}\nu_2^{-r_2/2}q_1^{-r_1/2}q_2^{-r_2/2} E \left\{Z_1^{r_1}\sum^{r_2}_{i=0}C(r_2,i)Z_1^i Z_2^{r_2-i} \sin^i(\theta)\cos^{r_2-i}(\theta)\right\}\\
&=&\nu_1^{-r_1/2}\nu_2^{-r_2/2}  q_1^{-r_1/2}q_2^{-r_2/2}  \sum^{r_2}_{i=0} \{ C(r_2,i) f(r_1+i)f(r_2-i) \sin^i(\theta)\cos^{r_2-i}(\theta)\}.
\end{eqnarray*}
Integrating over $q_1$ and $q_2$, we have
\begin{eqnarray*}
E(X_1^{r_1}X_2^{r_2})&=& (2\nu_1)^{-r_1/2}(2\nu_2)^{-r_2/2}  \frac{\Gamma \left( \frac{\nu_1-r_1}{2}\right)\Gamma \left( \frac{\nu_2-r_2}{2}\right)}{\Gamma \left( \frac{\nu_1}{2}\right)\Gamma \left( \frac{\nu_2}{2}\right)} \\
&& \cdot  \sum^{r_2}_{i=0} \{ C(r_2,i) f(r_1+i)f(r_2-i) \sin^i(\theta)\cos^{r_2-i}(\theta)\}
\end{eqnarray*}
for $\nu_1 >r_1$ and $\nu_2 >r_2$.
For $p>2$, the product moment expectation can be derived following the same procedure.

\citet{shaw2008bivariate} derived the explicit form  of the density function for bivariate $t$ distribution with variable marginal numbers of degrees of freedom and independence, which is actually a special case of the NECTD.
The density of our bivariate NECTD is 
\begin{eqnarray*}
&&f(x_1,x_2)\\
&=&C \alpha_1^{-\nu_1/2-1} \alpha_2^{-\nu_2/2-1}\Bigg \{ {}_2F_1 \left ( \frac{\nu_1+1}{2},  \frac{\nu_2+1}{2}; \frac{1}{2}; \frac{\gamma^2}{4 \alpha_1 \alpha_2 }\right)  \Gamma \left( \frac{\nu_1+1}{2}\right) \Gamma \left( \frac{\nu_2+1}{2}\right)    \\
&&\cdot \sqrt{\alpha_1 \alpha_2}  
+ {}_2 F_1 \left ( \frac{\nu_1}{2}+1,  \frac{\nu_2}{2}+1; \frac{3}{2}; \frac{\gamma^2}{4 \alpha_1 \alpha_2 }\right)  \gamma \Gamma \left( \frac{\nu_1}{2}+1\right) \Gamma \left( \frac{\nu_2}{2}+1\right)  \Bigg \},
\end{eqnarray*}
where
\begin{eqnarray*}
&&\alpha_1=1+\frac{x_1^2}{\nu_1 \cos^2(\theta)}, \quad  \alpha_1=1+\frac{x_2^2}{\nu_2 \cos^2(\theta)}, \quad \gamma=\frac{2x_1x_2\sin(\theta)}{\sqrt{\nu_1 \nu_2} \cos^2(\theta)},\\
&&C=\frac{1}{\cos(\theta)\pi \sqrt{\nu_1\nu_2}\Gamma(\nu_1/2)\Gamma(\nu_2/2)},
\end{eqnarray*}
and ${}_2 F_1(\cdot)$ is the hypergeometric function.
For $p>2$, it is too complicate to give the form of the density.

\section*{Appendix B: Bayesian Inference for NECTD}
We present technical details of  Bayesian computation for the NECTD.  Based on (3), we
treat $\{ \bm{Q}_i: i=1,\ldots,m\}$ as missing data, and write the likelihood for the complete data as
\begin{eqnarray*}
&&\prod_{i=1}^n   \Big|   \bm{Q}_i^{ - 1/2} \bm{\Sigma} \bm{Q}_i^{- 1/2} \Big|  ^{-1/2}   \exp\left\{      -\frac{1}{2} (\bm{X}_i - \bm{\mu})^\top \bm{Q}_i^{1/2} \bm{\Sigma}^{-1} \bm{Q}_i^{1/2} (\bm{X}_i - \bm{\mu})    \right\} \\
&&\cdot \prod_{i=1}^n \prod_{j=1}^p  \left( \frac{ 2}{ \nu_j } \right)^{\nu_j/2} \Gamma^{-1}  \left(  \frac{\nu_j }{ 2 } \right) q_{ij}^{\nu_j/2 - 1} e^{-q_{ij} \nu_j / 2}\\
&\propto&   |   \bm{\Sigma}   |^{-n/2}   \exp\left\{          -\frac{1}{2} \sum_{i=1}^n (\bm{X}_i - \bm{\mu})^\top \bm{Q}_i^{1/2} \bm{\Sigma}^{-1} \bm{Q}_i^{1/2} (\bm{X}_i - \bm{\mu})       \right\}\\
&&\cdot \prod_{i=1}^n \prod_{j=1}^p  \left( \frac{ \nu_j }{ 2} \right)^{\nu_j/2} \Gamma^{-1}  \left(  \frac{\nu_j }{ 2 } \right) q_{ij}^{(\nu_j - 1) /2 } e^{-q_{ij} \nu_j / 2}.
\end{eqnarray*}
For Bayesian inference, we need to specify prior distributions for all the parameters $(\bm{\mu},\bm{\Sigma},\bm{\nu})$. We choose a multivariate Normal prior for the mean vector, $\bm{\mu}\sim \bm{N}_p(\bm{\mu}_0, \bm{\Sigma}_0)$, an inverse-Wishart prior for the scale matrix, $\bm{\Sigma}\sim \text{Inv-Wishart}(\nu_0, \bm{I}_p)$, and Gamma priors for the numbers of degrees of freedom,
$\nu_j\sim \text{Gamma}(\theta_0, \phi_0)$.

\subsection*{Imputation Step}
First, we  impute all the missing $\bm{Q}_i$'s. 
The posterior density of $q_{ij}$ is
\begin{eqnarray*}
f(q_{ij}\mid \cdot) &\propto& \exp\left\{     -\frac{1}{2} (\bm{X}_i - \bm{\mu})^\top \bm{Q}_i^{1/2} \bm{\Sigma}^{-1} \bm{Q}_i^{1/2} (\bm{X}_i - \bm{\mu})  + \frac{\nu_j - 1}{2} \log q_{ij}  - \frac{\nu_j}{2}  q_{ij}  \right\},
\end{eqnarray*}
where 
\begin{eqnarray*}
 &&(\bm{X}_i - \bm{\mu})^\top \bm{Q}_i^{1/2} \bm{\Sigma}^{-1} \bm{Q}_i^{1/2} (\bm{X}_i - \bm{\mu}) \\
 &=& ( X_{i1} - \mu_1  , \cdots, X_{ip} - \mu_p)
 \begin{pmatrix}
 q_{i1} \sigma^{11}  &  \sqrt{q_{i1}q_{i2}} \sigma^{12} &  \cdots & \sqrt{q_{i1} q_{ip}} \sigma^{1p} \\
  \sqrt{q_{i2}q_{i1}} \sigma^{21} &  q_{i2} \sigma^{22} &  \cdots & \sqrt{q_{i2} q_{ip}} \sigma^{2p} \\
 \vdots & \vdots &  \cdots & \vdots \\
 \sqrt{q_{ip}q_{i1}} \sigma^{p1} &  \sqrt{q_{ip}q_{i2}} \sigma^{p2} &  \cdots & \sqrt{q_{ip} q_{ip}} \sigma^{pp}
 \end{pmatrix}
 \begin{pmatrix}
 X_{i1} - \mu_1 \\
 \vdots\\
 X_{ip} - \mu_p
 \end{pmatrix}  \\
 &=& q_{ij} \sigma^{jj} (X_{ij} - \mu_j)^2
 +
 2\sqrt{q_{ij}} (X_{ij} - \mu_j)\sum_{k\neq j} \sqrt{q_{ik}} \sigma^{jk} (X_{ik} - \mu_k) +C_{ij},
\end{eqnarray*}
and $C_{ij}$ is  a constant independent of $q_{ij}$. Thus, 
\begin{eqnarray}
\nonumber &&f(q_{ij}\mid \cdot) \\
\nonumber &\propto& \exp\left[      -\frac{ q_{ij}   }{2}  \left\{  \nu_j + \sigma^{jj} (X_{ij} - \mu_j)^2 \right\}
- \sqrt{q_{ij}}  (X_{ij} - \mu_j)\sum_{k\neq j} \sqrt{q_{ik}} \sigma^{jk} (X_{ik} - \mu_k)
 +  \frac{\nu_j - 1}{2} \log q_{ij}
\right]
\\
\label{eqn:MISnu}&=&   \exp\left(     -\frac{ q_{ij}   }{2} u_{ij}
- \sqrt{q_{ij}} c_{ij}
 +  \frac{\nu_j - 1}{2} \log q_{ij}
\right) ,
\end{eqnarray}
where $u_{ij} =   \nu_j + \sigma^{jj} (X_{ij} - \mu_j)^2 > 0,$ and $c_{ij} =  (X_{ij} - \mu_j)\sum_{k\neq j} \sqrt{q_{ik}} \sigma^{jk} (X_{ik} - \mu_k) . $

The posterior distribution of $q_{ij}$ is not standard, and we propose a Metropolized Independence Sampler (MIS) to  sample $q_{ij}$ based on a Gamma approximation  \citep{liu2008monte}.
 The MIS is a special case of the Metropolis-Hastings algorithm. In each step of the MCMC, instead of generating a candidate sample dependent of the previous sample, we independently generate a candidate sample  $q_{ij}^*$ from a Gamma distribution,  $q_{ij}^* \sim \text{Gamma}(\alpha, \beta)$. After generating the candidate sample, we can calculate the acceptance rate, and then decide whether to accept it.

If $ \nu_j \leq 1 $, then $f(q_{ij}\mid \cdot)$ is decreasing in $q_{ij}$, and we then choose an exponential distribution with $\alpha=1$ and $\beta=u_{ij}/2$ .

If $ \nu_j >1 $,
we choose $\alpha$ and $\beta$ to match the mode and the curvature at mode of the Gamma proposal with those of $f(q_{ij}\mid \cdot)$.  The mode of $\text{Gamma}(\alpha, \beta)$ is $\left( { \alpha-1\over\beta}\right)$ and the curvature at mode is $\left( - {\beta^2 \over \alpha - 1} \right)$.
%we match the mode $\left( { \alpha-1\over\beta}\right)$ and the curvature at mode $\left( - {\beta^2 \over \alpha - 1} \right)$ of the Gamma proposal with that of $f(q_{ij}\mid \cdot)$.
Denote the mode and the curvature at mode of $f(q_{ij}\mid \cdot)$ at the mode by $m_{ij}^*$ and $l_{ij}^*$, respectively.  By solving 
\begin{eqnarray*}
\frac{\alpha-1}{\beta}=m_{ij}^*, \quad -\frac{\beta^2}{\alpha-1}=l_{ij}^*,
\end{eqnarray*}
we have $\alpha = 1 - l_{ij}^* m_{ij}^{*2}$ and $ \beta = -l_{ij}^* m_{ij}^*$.
From \eqref{eqn:MISnu}, by solving
\begin{eqnarray*}
{ \partial \log f(q_{ij}\mid \cdot) \over \partial q_{ij}  } &=&   -\frac{u_{ij} }{2}  - \frac{c_{ij}}{2\sqrt{q_{ij}}} +    \frac{\nu_j - 1}{2q_{ij}} =0,
%{ \partial ^2 \log f(q_{ij}\mid \cdot) \over \partial q_{ij}^2  } &=&
%\frac{c_{ij}}{4\sqrt{q_{ij}^3}} - \frac{\nu_j - 1}{2q_{ij}^2}.
\end{eqnarray*}
we have 
$$
m_{ij}^* = \left(  \frac{ c_{ij}/2 + \sqrt{  (c_{ij}/2)^2 + u_{ij}(\nu_j - 1)  }  }{\nu_j - 1}      \right)^{-2}.
$$
Then, 
$$
 l_{ij}^*=\left. { \partial ^2 \log f(q_{ij}\mid \cdot) \over \partial q_{ij}^2  } \right |_{q_{ij}=m_{ij}^*} =\frac{c_{ij}}{4\sqrt{m_{ij}^{*3}}} - \frac{\nu_j - 1}{2m_{ij}^{*2}}.
$$
%If $\nu_j >1 $, we can calculate that 
%
%We choose the shape parameter $\alpha$ and the rate parameter $\beta$ of the Gamma distribution as 
%$$
%\alpha = 1 - l_{ij}^* m_{ij}^{*2} \text{ and } \beta = -l_{ij}^* m_{ij}^*.
%$$
After obtaining $\alpha$ and $\beta$ for the Gamma proposal,
we generate a candidate sample from $\text{Gamma}(\alpha, \beta)$ and then calculate the acceptance rate.  By generating a binary random variable, we can decide whether to accept the candidate sample.

\subsection*{Posterior Step}
The posterior distribution is proportional to
\begin{eqnarray*}
&&   |   \bm{\Sigma}   |^{-n/2}   \exp\left\{          -\frac{1}{2} \sum_{i=1}^n (\bm{X}_i - \bm{\mu})^\top \bm{Q}_i^{1/2} \bm{\Sigma}^{-1} \bm{Q}_i^{1/2} (\bm{X}_i - \bm{\mu})       \right\}\\
&&\cdot \prod_{i=1}^n \prod_{j=1}^p  \left( \frac{ \nu_j }{ 2} \right)^{\nu_j/2} \Gamma^{-1}  \left(  \frac{\nu_j }{ 2 } \right) q_{ij}^{(\nu_j - 1) /2 } e^{-q_{ij} \nu_j / 2} \\
&& \cdot \exp\left\{   -\frac{1}{2}(\bm{\mu}-\bm{\mu_0})^\top \bm{\Sigma}_0^{-1} (\bm{\mu}-\bm{\mu_0})   \right\}  \cdot  |   \bm{\Sigma}   |^{-\frac{\nu_0+p+1}{2}} \exp \left\{- \frac{1}{2} \text{tr}(\bm{\Sigma}^{-1}) \right\}   \cdot \prod_{j=1}^p    \nu_j^{\theta_0-1}  e^{-\phi_0 \nu_j}.
\end{eqnarray*}
From the  decomposition:
\begin{eqnarray*}
&&    \sum_{i=1}^n (\bm{X}_i - \bm{\mu})^\top \bm{Q}_i^{1/2} \bm{\Sigma}^{-1} \bm{Q}_i^{1/2} (\bm{X}_i - \bm{\mu})  +  (\bm{\mu}-\bm{\mu_0})^\top \Sigma_0^{-1} (\bm{\mu}-\bm{\mu_0})  \\
&=&   \sum_{i=1}^n (\bm{X}_i - \widehat{\bm{\mu}})^\top \bm{Q}_i^{1/2} \bm{\Sigma}^{-1} \bm{Q}_i^{1/2} (\bm{X}_i - \widehat{\bm{\mu}})  +  (\widehat{\bm{\mu}}-\bm{\mu_0})^\top \Sigma_0^{-1} (\widehat{\bm{\mu}}-\bm{\mu_0}) \\
&&+ (\bm{\mu} - \widehat{\bm{\mu}})^\top \left( \sum_{i=1}^n \bm{Q}_i^{1/2} \bm{\Sigma}^{-1} \bm{Q}_i^{1/2}
+ \bm{\Sigma}_0^{-1} \right) ( \bm{\mu} - \widehat{\bm{\mu}}),
\end{eqnarray*}
where
\begin{eqnarray*}
&&  \widehat{\bm{\mu}} = \left( \sum_{i=1}^n \bm{Q}_i^{1/2} \bm{\Sigma}^{-1} \bm{Q}_i^{1/2}
+ \bm{\Sigma}_0^{-1} \right)^{-1} \left( \sum_{i=1}^n \bm{Q}_i^{1/2} \bm{\Sigma}^{-1} \bm{Q}_i^{1/2}\bm{X}_i
+ \bm{\Sigma}_0^{-1}\bm{\mu_0} \right),
\end{eqnarray*}
we obtain
the conditional posterior density of $\bm{\mu}$:
\begin{eqnarray*}
\bm{\mu}|\cdot \sim  \bm{N}_p\left\{\widehat{\bm{\mu}},\left(\sum^{n}_{i=1}\bm{Q}_i^{1/2} \bm{\Sigma}^{-1} \bm{Q}_i^{1/2}+\bm{\Sigma}_0^{-1}\right)^{-1}\right\}.
\end{eqnarray*}
Then, the conditional posterior density of $\bm{\Sigma}$ is
\begin{eqnarray*}
\bm{\Sigma}| \cdot 
&\sim&  \text{Inv-Wishart}\left\{ n+\nu_0, \bm{I}_p+ \sum^{n}_{i=1}\bm{Q}_i^{1/2} (\bm{X}_i-\bm{\mu})(\bm{X}_i-\bm{\mu})^\top \bm{Q}_i^{1/2} \right\}.
\end{eqnarray*}
The conditional posterior density of $\nu_j$ is
\begin{eqnarray*}
f(\nu_j| \cdot) &\propto& \exp \left\{ -\frac{n\nu_j}{2}\log \left(  \frac{\nu_j}{2} \right)
 - n\log\Gamma\left( \frac{\nu_j}{2} \right) + \frac{\nu_j}{2} \sumn \log q_{ij} -\frac{\nu_j}{2} \sumn q_{ij} \right.
 \\
 &&\left.+ (\theta_{0} - 1)\log \nu_j - \phi_{0} \nu_j \right\}\\
 &\propto& \exp \left\{ -\frac{n\nu_j}{2}\log \left(  \frac{\nu_j}{2} \right)
 - n\log\Gamma\left( \frac{\nu_j}{2} \right) + (\theta_{0} - 1)\log \nu_j - \eta_{j} \nu_j \right\},
\end{eqnarray*}
where $\eta_j = \phi_{0} - \sum_{i=1}^{n} \log q_{ij}/2 + \sum_{i=1}^{n} q_{ij}/2.$
To sample $\nu_j$, we can also use the MIS based on a Gamma approximation. The steps are similar to those of sampling $q_{ij}$.  In each step, we first generate a candidate sample from a Gamma distribution, then  calculate the acceptance rate, and finally decide whether to accept it. The mode and the curvature at mode of the Gamma distribution  are the same as those of the conditional distribution of $\nu_j$.

\section*{Appendix C: Generalized Selection-$t$ Model}
\renewcommand {\theequation} {C.\arabic{equation}}
\setcounter{equation}{0}
The outcome equation is
$$
y_i^* = \bm{x}_i^\top \bm{\beta} + \varepsilon_i,
$$
and the selection equation is
$$
u_i^* = \bm{w}_i^\top \bm{\gamma} + \eta_i.
$$
The error terms follow an NECTD:
$$
\begin{pmatrix}
\varepsilon_i\\
\eta_i
\end{pmatrix}
\sim \bm{t}_2\left\{ \begin{pmatrix}
0\\
0
\end{pmatrix},\bm{\Omega}, 
\bm{p} =
\begin{pmatrix}
1\\
1
\end{pmatrix},
\bm{\nu} = 
\begin{pmatrix}
\nu_1\\
\nu_2
\end{pmatrix} \right \}.
$$
We can represent the error terms  as
$$
\begin{pmatrix}
\varepsilon_i\\
\eta_i
\end{pmatrix}
=
\begin{pmatrix}
q_{i1}^{-1/2} & 0 \\
0& q_{i2}^{-1/2}
\end{pmatrix}
\bm{\Omega}^{1/2} \bm{Z}_i,
$$
where $q_{1i} \sim \chi^2_{\nu_1}/\nu_1,q_{1i} \sim \chi^2_{\nu_2}/\nu_2 ,  \bm{Z}_i\sim \bm{N}_2(\bm{0}_2, \bm{I}_2)$, and $(q_{1i}, q_{2i}, \bm{Z}_i)$ are jointly independent.

For the generalized selection-$t$ model, direct sampling the covariance matrix $\bm{\Omega}$ involves non-standard distributions. We solve this problem using parameter expansion. 
Denote
\begin{eqnarray*}
\bm{V}_i = \begin{pmatrix}
\bm{x}_i^\top & 0\\
0      &  \bm{w}_i^\top
\end{pmatrix},    \bm{Q}_i = \begin{pmatrix}
q_{i1} & 0 \\
0& q_{i2}
\end{pmatrix},  
   \bm{Z}_i =\begin{pmatrix}
y_i^* \\
u_i^*
\end{pmatrix},  \bm{\Sigma} =\begin{pmatrix}
1 & 0 \\
0& \sigma_2
\end{pmatrix},
\bm{\Omega} \begin{pmatrix}
1 & 0 \\
0& \sigma_2
\end{pmatrix},  
\bm{\delta}= \begin{pmatrix}
\bm{\beta} \\
\bm{\gamma}
\end{pmatrix}.
\end{eqnarray*}

We choose a multivariate Normal prior for the regression coefficients, $\bm{\delta} \sim \bm{N}_{K+L}(\bm{\delta}_0,\bm{\Sigma}_0)$, an inverse-Wishart prior for the covariance matrix, $\bm{\Sigma} \sim \text{Inv-Wishart}(\nu_0,\bm{I}_2)$, and   Gamma priors for the numbers of degrees of freedom,
$\nu_j\sim \text{Gamma}(\theta_0, \phi_0)$.

The prior for $\bm{\Sigma}$ is equivalent to
\begin{eqnarray}
f(\bm{\Omega}) &\propto& (1-\rho^2)^{-3/2}\sigma_1^{-\nu_0+3}\exp{\left\{-\frac{1}{2\sigma_1^2(1-\rho^2)}\right\}}, \label{selt:prior1} \\
 \sigma_2^2 \mid \bm{\Omega} & \sim& \{(1-\rho^2) \chi^2_{\nu_0}\}^{-1}.\label{selt:prior2} 
\end{eqnarray}

The complete-data likelihood is
\begin{eqnarray*}
&&\prod_{i=1}^n   \Big|   \bm{Q}_i^{ - 1/2} \bm{\Omega} \bm{Q}_i^{- 1/2} \Big|  ^{-1/2}   \exp\left\{      -\frac{1}{2} (\bm{Z}_i - \bm{V}_i\bm{\delta})^\top \bm{Q}_i^{1/2} \bm{\Omega}^{-1} \bm{Q}_i^{1/2} (\bm{Z}_i - \bm{V}_i\bm{\delta})    \right\} \\
&&\cdot \prod_{i=1}^n \prod_{j=1}^2  \left( \frac{ \nu_j }{ 2} \right)^{\nu_j/2} \Gamma^{-1}  \left(  \frac{\nu_j }{ 2 } \right) q_{ij}^{\nu_j/2 - 1} e^{-q_{ij} \nu_j / 2}\\
&\propto&   |   \bm{\Omega}   |^{-n/2}   \exp\left\{          -\frac{1}{2} \sum_{i=1}^n (\bm{Z}_i - \bm{V}_i\bm{\delta})^\top \bm{Q}_i^{1/2} \bm{\Omega}^{-1} \bm{Q}_i^{1/2} (\bm{Z}_i - \bm{V}_i\bm{\delta})       \right\}\\
&&\cdot \prod_{i=1}^n \prod_{j=1}^2  \left( \frac{ \nu_j }{ 2} \right)^{\nu_j/2} \Gamma^{-1}  \left(  \frac{\nu_j }{ 2 } \right) q_{ij}^{(\nu_j - 1) /2 } e^{-q_{ij} \nu_j / 2}.
\end{eqnarray*}

\subsection*{Imputation Step}
First, we impute the missing data given the observed data and the parameters.
Let $TN(\mu,\sigma^2;L,U)$ be a Normal distribution $N(\mu,\sigma^2)$ truncated within the interval $[L,U]$.
Given $(q_{i1},q_{i2},y_i,u_i,\bm{\Omega},\nu_1,\nu_2,\bm{\delta})$, we impute $(y_i^*,u_i^*)$ as follows: if $u_i=1$, we draw $y_i^* =y_i$ and
\begin{eqnarray*}
%y_i^* &= &y_i, \\
u_i^* \mid (y_i^*, q_{i1},q_{i2},y_i,u_i,\bm{\Omega},\nu_1,\nu_2,\bm{\delta}) &\sim  &  TN(\mu_{u\mid y},\sigma^2_{u|y};0,\infty),
\end{eqnarray*}
where
\begin{eqnarray*}
\mu_{u\mid y} = \bm{w}_i^\top \bm{\gamma}+\sqrt{\frac{q_{i1}}{q_{i2}}}\frac{\rho(y_i^*-\bm{x}_i^\top\bm{\beta})}{\sigma_1}, \quad   &&
\sigma^2_{u\mid y} = \frac{1-\rho^2}{q_{i2}};
\end{eqnarray*}
if $u_i=0$, we draw
\begin{eqnarray*}
&u_i^* \mid( q_{i1},q_{i2},y_i,u_i,\bm{\Omega},\nu_1,\nu_2,\bm{\delta}) &\sim \quad TN(\bm{w}_i^\top \bm{\gamma},1/q_{i2};-\infty,0), \\
&y_i^* \mid (u_i^*, q_{i1},q_{i2},y_i,u_i,\bm{\Omega},\nu_1,\nu_2,\bm{\delta}) &\sim \quad   N(\mu_{y\mid u},\sigma^2_{y|u}),
\end{eqnarray*}
where
\begin{eqnarray*}
\mu_{y\mid u} = \bm{x}_i^\top\bm{\beta}+\sqrt{\frac{q_{i2}}{q_{i1}}}\rho\sigma_1(u_i^*-\bm{w}_i^\top \bm{\gamma}), \quad &&
\sigma^2_{y\mid u} = \frac{\sigma_1^2(1-\rho^2)}{q_{i1}}.
\end{eqnarray*}

Denote $\bm{\Omega}^{-1}=\{\omega^{kl}\}$. Given  $(y_i^*,u_i^*,\bm{\Omega},\nu_1,\nu_2,\bm{\delta})$, we draw
%can impute $(q_{i1},q_{i2})$ as follows:
\begin{eqnarray*}
&q_{i1} \mid (q_{i2},y_i^*,u_i^*,\bm{\Omega},\nu_1,\nu_2,\bm{\delta}) \propto \exp{\left( -\frac{u_{i1}}{2}q_{i1}-c_{i1}\sqrt{q_{i1}}+\frac{\nu_1-1}{2}\log{q_{i1}}\right)},\\
%\text{and }
&q_{i2} \mid (q_{i1},y_i^*,u_i^*,\bm{\Omega},\nu_1,\nu_2,\bm{\delta}) \propto \exp{\left( -\frac{u_{i2}}{2}q_{i2}-c_{i2}\sqrt{q_{i2}}+\frac{\nu_1-1}{2}\log{q_{i2}}\right)},
\end{eqnarray*}
where
\begin{eqnarray*}
u_{i1}=\nu_1+\omega^{11}(y_i^*-\bm{x}_i^\top\bm{\beta})^2, && c_{i1}=\sqrt{q_{i2}}\omega^{12}(y_i^*-\bm{x}_i^\top\bm{\beta})(u_i^*-\bm{w}_i^\top \bm{\gamma}),\\
u_{i2}=\nu_2+\omega^{22}(u_i^*-\bm{w}_i^\top \bm{\gamma})^2, && c_{i2}=\sqrt{q_{i1}}\omega^{12}(y_i^*-\bm{x}_i^\top\bm{\beta})(u_i^*-\bm{w}_i^\top \bm{\gamma}).
\end{eqnarray*}

\subsection*{Posterior Step}
The posterior distribution is proportional to
\begin{eqnarray*}
&&    |   \bm{\Omega}   |^{-n/2}   \exp\left\{          -\frac{1}{2} \sum_{i=1}^n (\bm{Z}_i - \bm{V}_i\bm{\delta})^\top \bm{Q}_i^{1/2} \bm{\Omega}^{-1} \bm{Q}_i^{1/2} (\bm{Z}_i - \bm{V}_i\bm{\delta})       \right\}\\
&&\cdot \prod_{i=1}^n \prod_{j=1}^2  \left( \frac{ \nu_j }{ 2} \right)^{\nu_j/2} \Gamma^{-1}  \left(  \frac{\nu_j }{ 2 } \right) q_{ij}^{(\nu_j - 1) /2 } e^{-q_{ij} \nu_j / 2} \\
&& \cdot \exp\left\{   -\frac{1}{2}(\bm{\delta}-\bm{\delta}_0)^\top \bm{\Sigma}_0^{-1} (\bm{\delta}-\bm{\delta}_0)   \right\}  \cdot  (1-\rho^2)^{-3/2}\sigma_1^{-\nu_0+3}\exp{\left\{-\frac{1}{2\sigma_1^2(1-\rho^2)}\right\}}  \\
&& \cdot \prod_{j=1}^2    \nu_j^{\theta_0-1}  e^{-\phi_0 \nu_j}.
\end{eqnarray*}
We  draw
$
\bm{\delta}\mid \{y_i^*,u_i^*,q_{i1},q_{i2},\bm{\Omega},\nu_1,\nu_2\} \sim \bm{N}_{K+L}(\widehat{\bm{\mu}}_{\delta},\widehat{\bm{\Sigma}}_{\delta}),
$
where
\begin{eqnarray*}
 \widehat{\bm{\mu}}_{\delta} =  \widehat{\bm{\Sigma}}_{\delta} \left( \sum_{i=1}^n\bm{V}_i^\top \bm{Q}_i^{1/2} \bm{\Sigma}^{-1} \bm{Q}_i^{1/2}\bm{Z}_i
+ \bm{\Sigma}_0^{-1}\bm{\delta}_0 \right),  \quad
\widehat{\bm{\Sigma}}_{\delta}=\left( \sum_{i=1}^n \bm{V}_i^\top\bm{Q}_i^{1/2} \bm{\Omega}^{-1} \bm{Q}_i^{1/2}\bm{V}_i
+ \bm{\Sigma}_0^{-1} \right)^{-1}.
\end{eqnarray*}

To draw $\bm{\Omega}$, we use parameter expansion to re-parametrize the model and get a conjugate posterior distirbution. Define
\begin{eqnarray}
\bm{E}_i = \begin{pmatrix}
1 &0\\
0 & \sigma_2
\end{pmatrix} (\bm{Z}_i-\bm{V}_i\bm{\delta}), \label{selt:trans}
\end{eqnarray}
and we have $\bm{E}_i| (\bm{Q}_i, \bm{\delta},\bm{\nu},\sigma_2)\sim \bm{N}_2(\bm{0}_2, \bm{Q}_i^{-1/2} \bm{\Sigma} \bm{Q}_i^{-1/2})$. Because the prior of $\bm{\Sigma}$ implies priors in (\ref{selt:prior1}) and (\ref{selt:prior2}), we first draw $\sigma_2^{2}| \bm{\Omega} \sim \{(1-\rho^2) \chi^2_{\nu_0}\}^{-1}$, and then transform the data to get $\bm{E}_i$ using (\ref{selt:trans}).
The conditional posterior of $\bm{\Sigma}$ is $\text{Inv-Wishart}(n+\nu_0, E+\bm{I}_2)$, where $E=\sum_{i=1}^n \bm{Q}_i^{1/2} \bm{E}_i \bm{E}_i^\top \bm{Q}_i^{1/2}$. After drawing $\bm{\Sigma}$,  we transform $\bm{\Sigma}$ to
\begin{eqnarray*}
\sigma_2^2 = \sigma_{22} 
\text{ and } 
\bm{\Omega} =\begin{pmatrix}
1&0\\
0&1/\sigma_2
\end{pmatrix}  \bm{\Sigma}
\begin{pmatrix}
1&0\\
0&1/\sigma_2
\end{pmatrix}.
\end{eqnarray*}

Given $(y_i^*,u_i^*,q_{i1},q_{i2},\bm{\Omega},\bm{\delta})$, the conditional posterior density of $\nu_j$ is
\begin{eqnarray*}
\nu_j\mid (y_i^*,u_i^*,q_{i1},q_{i2},\bm{\Omega},\bm{\delta}) \propto \exp \left\{ -\frac{n\nu_j}{2}\log \left(  \frac{\nu_j}{2} \right)
 - n\log\Gamma\left( \frac{\nu_j}{2} \right) + (\theta_{0} - 1)\log \nu_j - \eta_{j} \nu_j \right\},
\end{eqnarray*}
where  $\eta_j = \phi_{0} -\sum_{i=1}^{n} \log q_{ij}/2 + \sum_{i=1}^{n} q_{ij}/2$. Following the same steps of drawing $\nu_j$ in the Bayesian inference for NECTD, we use the MIS based on a Gamma approximation to draw $\nu_j$.

\section*{Appendix D: Generalized Robit Model}
\renewcommand {\theequation} {D.\arabic{equation}}
\setcounter{equation}{0}
The observed variables $\bm{y}_i=(y_{i1},\ldots, y_{ip})^\top$ are truncated versions of latent variables $\bm{y}^*_i=(y^*_{i1},\ldots, y^*_{ip})^\top$ via $y_{ij}=I(y^*_{ij}>0)$:
%, with the latent variables modeled as 
\begin{eqnarray*}
\bm{y}^*_i = \bm{x}_i \bm{\beta}+\bm{\varepsilon_i}, \quad \bm{\varepsilon_i} \sim \bm{t}_p(\bm{0}_p, \bm{\Omega}, \bm{p}, \bm{\nu}),
\end{eqnarray*}
 where  $\bm{x}_i$  is a known $p\times K$  design matrix. 

Because direct sampling of the covariance matrix involves non-standard distributions, we solve this problem using parameter expansion. Denote 
$$
\bm{\Sigma}=\text{diag}\{  d_1,\ldots,d_p\} ~ \bm{\Omega}~\text{diag}\{ d_1,\ldots,d_p\}.
$$
We choose a multivariate Normal prior for the regression coefficients, $\bm{\beta} \sim \bm{N}_{k}(\bm{\beta}_0,\bm{\Sigma}_0)$, an inverse-Wishart prior for the covariance matrix, $\bm{\Sigma} \sim \text{Inv-Wishart}(\nu_0,\bm{I}_p)$, and  Gamma priors for the numbers of degrees of freedom,
$\nu_j\sim \text{Gamma}(\theta_0, \phi_0)$. 
The prior for $\bm{\Sigma}$ is equivalent to
\begin{eqnarray}
&f(\bm{\Omega}) &\propto \quad | \bm{\Omega} |^{-(\nu_0+p+1)/2} \left(\prod_i \omega^{ii}\right)^{-\nu_0/2}, \label{gr:prior1} \\
&d_i^2 \mid \bm{\Omega} & \sim \quad  \omega^{ii}/\chi^2_{\nu_0}. \label{gr:prior2}
\end{eqnarray}
The complete-data likelihood is
\begin{eqnarray*}
&&\prod_{i=1}^n   \Big|   \bm{Q}_i^{ - 1/2} \bm{\Omega} \bm{Q}_i^{- 1/2} \Big|  ^{-1/2}   \exp\left\{      -\frac{1}{2} (\bm{y}^*_i - \bm{x}_i\bm{\beta})^\top \bm{Q}_i^{1/2} \bm{\Omega}^{-1} \bm{Q}_i^{1/2} (\bm{y}^*_i - \bm{x}_i\bm{\beta})    \right\} \\
&&\cdot \prod_{i=1}^n \prod_{j=1}^{p}  \left( \frac{ \nu_j }{ 2} \right)^{\nu_j/2} \Gamma^{-1}  \left(  \frac{\nu_j }{ 2 } \right) q_{ij}^{\nu_j/2 - 1} e^{-q_{ij} \nu_j / 2} \\
&\propto&   |   \bm{\Omega}   |^{-n/2}   \exp\left\{          -\frac{1}{2} \sum_{i=1}^n (\bm{y}^*_i - \bm{x}_i\bm{\beta})^\top \bm{Q}_i^{1/2} \bm{\Omega}^{-1} \bm{Q}_i^{1/2} (\bm{y}^*_i - \bm{x}_i\bm{\beta})       \right\}\\
&&\cdot \prod_{i=1}^n \prod_{j=1}^{p}  \left( \frac{ \nu_j }{ 2} \right)^{\nu_j/2} \Gamma^{-1}  \left(  \frac{\nu_j }{ 2 } \right) q_{ij}^{(\nu_j - 1) /2 } e^{-q_{ij} \nu_j / 2}.
\end{eqnarray*}

\subsection*{Imputation Step}
Let $TN(\mu,\sigma^2;L,U)$ be a Normal distribution $N(\mu,\sigma^2)$ truncated within the interval $(L,U)$.
Given $(\bm{Q}_i,\bm{\beta}, \nu_j,\bm{\Omega},Y_i)$, we draw
\begin{eqnarray*}
W_{i,j} \sim TN(\mu_{ij},\sigma^2_{ij};L_{ij},U_{ij}),
\end{eqnarray*}
where $[ L_{ij},U_{ij} ] $ equals $ [0, +\infty]$  if $Y_{ij}=1$ and equals $[-\infty,0 ]$ if $ Y_{ij}=0$, and
\begin{eqnarray*}
\mu_{ij} &=& 
\bm{X}_{i,j} \bm{\beta}+q_{ij}^{-1/2} \bm{\Omega}_{j,-j} \bm{\Omega}_{-j,-j}^{-1}\{\bm{Q}_{i,-j}^{1/2}(\bm{y}^*_{i,-j}-\bm{x}_{i,-j} \bm{\beta})\},\\
\sigma_{ij}^2 &=& 
\frac{1}{q_{ij}}(\omega_{jj}- \bm{\Omega}_{j,-j} \bm{\Omega}_{-j,-j}^{-1}\bm{\Omega}_{-j,j}).
%\begin{cases}
%[0, +\infty]    & \text{if }   Y_{ij}=1, \\
%[-\infty,0 ]    & \text{if }   Y_{ij}=0. \\
%\end{cases}
\end{eqnarray*}

Given $(\bm{y}^*_i,\bm{\beta}, \nu_j,\bm{\Omega},Y_i)$, the conditional posterior density of  $q_{ij}$ is
\begin{eqnarray*}
f(q_{ij}\mid \cdot) &\propto&    \exp\left(      -\frac{ q_{ij}   }{2} u_{ij}
- \sqrt{q_{ij}} c_{ij}
 +  \frac{\nu_j - 1}{2} \log q_{ij}
\right) ,
\end{eqnarray*}
where 
$$
u_{ij} =   \nu_j + \omega^{jj} (y^*_{ij} - \bm{x}_{ij}\bm{\beta})^2 > 0,\quad
c_{ij} =  (y^*_{ij} - \bm{x}_{ij}\bm{\beta})\sum_{k\neq j} \sqrt{q_{ik}} \omega^{jk} (y^*_{ik} - \bm{x}_{ik}\bm{\beta}). 
$$ 
We draw  $q_{ij}$ using the same procedure as the imputation step of Appendix B.

\subsection*{Posterior Step}
The posterior distribution is proportional to
\begin{eqnarray*}
&&    |   \bm{\Omega}   |^{-n/2}   \exp\left\{          -\frac{1}{2} \sum_{i=1}^n (\bm{y}^*_i - \bm{x}_i\bm{\beta})^\top \bm{Q}_i^{1/2} \bm{\Omega}^{-1} \bm{Q}_i^{1/2} (\bm{y}^*_i - \bm{x}_i\bm{\beta})       \right\}\\
&&\cdot \prod_{i=1}^n \prod_{j=1}^{p}  \left( \frac{ \nu_j }{ 2} \right)^{\nu_j/2} \Gamma^{-1}  \left(  \frac{\nu_j }{ 2 } \right) q_{ij}^{(\nu_j - 1) /2 } e^{-q_{ij} \nu_j / 2} \\
&& \cdot \exp\left\{   -\frac{1}{2}(\bm{\beta}-\bm{\beta}_0)^\top \bm{\Sigma}_0^{-1} (\bm{\beta}-\bm{\beta}_0)   \right\}  \cdot | \bm{\Omega} |^{-(\nu_0+p+1)/2} \left (\prod_i \omega^{ii}\right)^{-\nu_0/2}\\
&&  \cdot \prod_{j=1}^{p}    \nu_j^{\theta_0-1}  e^{-\phi_0 \nu_j},
\end{eqnarray*}
where $\omega^{jj}$ is the $(j,j)$-th element of $\bm{\Omega}^{-1}$.

Given $(\bm{y}^*_i, \nu_j,\bm{\Omega},Y_i,\bm{Q}_i)$,  we draw
$
\bm{\beta}\mid \cdot \sim \bm{N}_{K}(\widehat{\bm{\mu}}_{\beta},\widehat{\bm{\Sigma}}_{\beta}),
$
where
\begin{eqnarray*}
 \widehat{\bm{\mu}}_{\beta} =  \widehat{\bm{\Sigma}}_{\beta} \left( \sum_{i=1}^n\bm{x}_i^\top \bm{Q}_i^{1/2} \bm{\Omega}^{-1} \bm{Q}_i^{1/2}\bm{y}^*_i
+ \bm{\Sigma}_0^{-1}\bm{\beta}_0 \right), \quad
\widehat{\bm{\Sigma}}_{\beta}=\left( \sum_{i=1}^n \bm{x}_i^\top\bm{Q}_i^{1/2} \bm{\Omega}^{-1} \bm{Q}_i^{1/2}\bm{x}_i
+ \bm{\Sigma}_0^{-1} \right)^{-1}.
\end{eqnarray*}

To draw $\bm{\Omega}$, we use parameter expansion. First, we draw $d_i^2 \mid \bm{\Omega}$ according to (\ref{gr:prior2}), then transform data to 
$
 \bm{E}_i = \bm{D}(\bm{y}^*_i-\bm{x}_i\bm{\beta}).
$
The conditional posterior density of $\bm{\Sigma}$ is $\text{Inv-Wishart}(n+\nu_0, E+\bm{I}_{p})$,
where $E=\sum_{i=1}^n \bm{Q}_i^{1/2} \bm{E}_i \bm{E}_i^\top \bm{Q}_i^{1/2}$. After drawing $\bm{\Sigma}$, we transform $\bm{\Sigma}$ to
\begin{eqnarray*}
d_i^2 = \Sigma_{ii}, \quad  \quad \bm{\Omega} = \bm{D}^{-1}\bm{\Sigma}\bm{D}^{-1}.
\end{eqnarray*}
Following the same steps of drawing $\nu_j$ in the Bayesian inference for NECTD,  
we use the MIS based on a Gamma approximation to draw
\begin{eqnarray*}
f(\nu_j\mid \bm{y}^*_i,\bm{\beta},\bm{\Omega},Y_i,\bm{Q}_i) \propto \exp \left\{ -\frac{n\nu_j}{2}\log \left(  \frac{\nu_j}{2} \right)
 - n\log\Gamma\left( \frac{\nu_j}{2} \right) + (\theta_{0} - 1)\log \nu_j - \eta_{j} \nu_j \right\},
\end{eqnarray*}
where  $\eta_j = \phi_{0} - \sum_{i=1}^{n} \log q_{ij}/2 + \sum_{i=1}^{n} q_{ij}/2.$

\section*{Appendix E: Generalized Linear $t$ Mixed-Effects Model}
For $i=1,\ldots,m$, the observed variables $\bm{y}_i=(y_{i1},\ldots,y_{in_i})^\top$ follow
\begin{eqnarray*}
\bm{y}_i=\bm{x}_i \bm{\beta}+\bm{z}_i \bm{b}_i+\bm{\epsilon}_i,
\end{eqnarray*}
where  $\bm{x}_i$ and $\bm{z}_i$ are known $n_i\times K$ and $n_i\times L$ design matrices corresponding to the $K$-dimensional fixed effects vector $\bm{\beta}$ and the $L$-dimensional random effects vector $\bm{b}_i$, respectively; $\bm{e}_i$ is an  $n_i$-dimensional vector error.
Assume 
 $$
\begin{pmatrix}
\bm{b}_i\\
\bm{\varepsilon}_i
\end{pmatrix}
\sim \bm{t}_{L+n_i}\left\{
\begin{pmatrix}
0\\
0
\end{pmatrix} ,
\begin{pmatrix}
 \bm{\Omega} & 0 \\
0& \bm{\Lambda}_i
\end{pmatrix}, 
\bm{p} =
\begin{pmatrix}
L\\
n_i
\end{pmatrix},
 \bm{\nu}=
 \begin{pmatrix}
 \nu_1\\
 \nu_2
 \end{pmatrix}
 \right\},
$$
where $\nu_1$ and $\nu_2$ are numbers of degrees of freedom for random effects and within-subject errors, respectively. 

We choose a multivariate Normal prior for the regression coefficients, $\bm{\beta} \sim \bm{N}_{k}(\bm{\beta}_0,\bm{\Sigma}_0)$, an inverse-Wishart prior for the covariance matrix, $\bm{\Sigma} \sim \text{Inv-Wishart}(\nu_0,\bm{I}_L)$, and   Gamma priors for the numbers of degrees of freedom,
$\nu_j\sim \text{Gamma}(\theta_0, \phi_0)$.  To guarantee a proper posterior distribution, we choose  $\sigma^2 \sim  \text{Inv-Gamma}(0.5,0.1)$ as the prior for the variance of the within-subject errors.

\subsection*{Imputation Step}
The complete-data likelihood is
\begin{eqnarray*}
&& \prod_{i=1}^{m} |\bm{\Omega} /q_{i1}|^{-1/2} \exp \left (-\frac{1}{2} q_{i1}\bm{b}_i^\top \bm{\Omega}^{-1}\bm{b}_i \right) \left( \frac{ \nu_1 }{ 2} \right)^{\nu_1/2} \Gamma^{-1}  \left(  \frac{\nu_1}{ 2 } \right) q_{i1}^{(\nu_1 - 2) /2 } e^{-q_{i1} \nu_1 / 2}\\
& \cdot& \prod_{i=1}^{m} (\sigma^2/q_{i2})^{-n_i/2} \exp  \left \{ -\frac{1}{2 \sigma^2} q_{i2}( \bm{y}_i-\bm{x}_i \bm{\beta}-\bm{z}_i \bm{b}_i)^\top ( \bm{y}_i-\bm{x}_i \bm{\beta}-\bm{z}_i \bm{b}_i)\right\}\\
&\cdot&\left( \frac{ \nu_2 }{ 2} \right)^{\nu_2/2} \Gamma^{-1}  \left(  \frac{\nu_2 }{ 2 } \right) q_{i2}^{(\nu_2 - 2) /2 } e^{-q_{i2} \nu_2 / 2}.
\end{eqnarray*}

Given $(\bm{b}_i, \bm{\Sigma}, \bm{\beta}, \sigma^2,\nu_1,\nu_2)$, we impute $q_{i1}$ and $q_{i2}$ from
\begin{eqnarray*}
q_{i1} \sim  \frac{\chi^2_{L+\nu_1}}{\bm{b}_i^\top \bm{\Omega}^{-1}\bm{b}_i+\nu_1}, \quad
\quad q_{i2} \sim  \frac{\chi^2_{n_i+\nu_2}}{( \bm{y}_i-\bm{x}_i \bm{\beta}-\bm{z}_i \bm{b}_i)^\top ( \bm{y}_i-\bm{x}_i \bm{\beta}-\bm{z}_i \bm{b}_i)/\sigma^2+\nu_2}.
\end{eqnarray*}

Given $(q_{i1},q_{i2}, \bm{\Sigma}, \bm{\beta}, \sigma^2,\nu_1,\nu_2)$, we impute $\bm{b}_i$ from
$
\bm{b}_i \sim \bm{N}_L (\widehat{\bm{\mu}}_b, \widehat{\bm{\Omega}}_b),
$
where
\begin{eqnarray*}
\widehat{\bm{\mu}}_b=\left( \frac{q_{i2}}{\sigma^2}\bm{z}_i^\top\bm{z}_i+q_{i1} \bm{\Omega}^{-1} \right )^{-1}  \frac{q_{i2}}{\sigma^2} \bm{z}_i^\top(\bm{y}_i-\bm{x}_i \bm{\beta}), \quad \widehat{\bm{\Omega}}_b=\left( \frac{q_{i2}}{\sigma^2}\bm{z}_i^\top\bm{z}_i+q_{i1} \bm{\Omega}^{-1} \right)^{-1}.
\end{eqnarray*}

\subsection*{Posterior Step}
 The posterior distribution is proportional to
 \begin{eqnarray*}
&& \prod_{i=1}^{m} |\bm{\Omega} /q_{i1}|^{-1/2} \exp \left ( -\frac{1}{2} q_{i1}\bm{b}_i^\top \bm{\Omega}^{-1}\bm{b}_i \right) \left( \frac{ \nu_1 }{ 2} \right)^{\nu_1/2} \Gamma^{-1}  \left(  \frac{\nu_1}{ 2 } \right) q_{i1}^{(\nu_1 - 2) /2 } e^{-q_{i1} \nu_1 / 2}\\
&& \cdot \prod_{i=1}^{m} (\sigma^2/q_{i2})^{-n_i/2} \exp  \left \{ -\frac{1}{2 \sigma^2} q_{i2}( \bm{y}_i-\bm{x}_i \bm{\beta}-\bm{z}_i \bm{b}_i)^\top ( \bm{y}_i-\bm{x}_i \bm{\beta}-\bm{z}_i \bm{b}_i)\right\}\\
&&\cdot \left( \frac{ \nu_2 }{ 2} \right)^{\nu_2/2} \Gamma^{-1}  \left(  \frac{\nu_2 }{ 2 } \right) q_{i2}^{(\nu_2 - 2) /2 } e^{-q_{i2} \nu_2 / 2}\\
&& \cdot (\sigma^2)^{-1.5} \exp \left(-\frac{0.1}{\sigma^2}\right)  \exp\left\{   -\frac{1}{2}(\bm{\beta}-\bm{\beta}_0)^\top \bm{\Omega}_0^{-1} (\bm{\beta}-\bm{\beta}_0)   \right\}  \cdot |\bm{\Omega}|^{-\frac{\nu_0+L+1}{2}} \exp \left\{ \frac{1}{2} \text{tr}(\bm{\Omega}^{-1})\right\} \\
&&  \cdot \prod_{j=1}^{2}    \nu_j^{\theta_0-1}  e^{-\phi_0 \nu_j}.
\end{eqnarray*}
 
 Given $\{\bm{b}_i, \bm{\beta}, q_{i1},q_{i2},\nu_1,\nu_2\}$, we draw $(\bm{\Omega},\sigma^2)$ from
 \begin{eqnarray*}
 &\bm{\Omega} &\sim \quad \text{Inv-Wishart} \left(\nu_0+m, \bm{I}_L+\sum_{i=1}^m q_{i1}\bm{b}_i \bm{b}_i^\top  \right),\\
 & \sigma^2 &\sim \quad   \frac{0.2+ \sum_{i=1}^m q_{i2}( \bm{y}_i-\bm{x}_i \bm{\beta}-\bm{z}_i \bm{b}_i)^\top ( \bm{y}_i-\bm{x}_i \bm{\beta}-\bm{z}_i \bm{b}_i)} {\chi^2_{\sum_{i=1}^m n_i+0.5}}.
\end{eqnarray*}
 Given $(\bm{b}_i, \bm{\Omega}, \sigma^2, q_{i1},q_{i2},\nu_1,\nu_2 )$, we draw $\bm{\beta}$ from
$
\bm{\beta} \sim \bm{N}_L (\widehat{\bm{\mu}}_{\beta}, \widehat{\bm{\Omega}}_{\beta}),
$
where 
\begin{eqnarray*}
\widehat{\bm{\Omega}}_{\beta}=\left( \sum_{i=1}^m \frac{q_{i2}}{\sigma^2}\bm{x}_i^\top\bm{x}_i+ \bm{\Omega}_0^{-1} \right)^{-1}, \quad
\widehat{\bm{\mu}}_{\beta}= \widehat{\bm{\Omega}}_{\beta}  \left \{ \frac{q_{i2}}{\sigma^2} \bm{x}_i^\top(\bm{y}_i-\bm{z}_i \bm{b}_i) +  \bm{\Omega}_0^{-1} \bm{\beta}_0 \right \}.
\end{eqnarray*}

 Given $(\bm{b}_i, \bm{\beta}, q_{i1},q_{i2}, \bm{\Omega},\sigma^2)$, we use  the MIS based on a Gamma approximation to draw $(\nu_1,\nu_2)$ from
 \begin{eqnarray*}
f(\nu_j \mid \cdot) \propto \exp \left\{ -\frac{m\nu_j}{2}\log \left(  \frac{\nu_j}{2} \right)
 - m\log\Gamma\left( \frac{\nu_j}{2} \right) + (\theta_{0} - 1)\log \nu_j - \eta_{j} \nu_j \right\},
\end{eqnarray*}
where  $\eta_j = \phi_{0} - \sum_{i=1}^{m} \log q_{ij}/2 + \sum_{i=1}^{m} q_{ij}/2.$

\section*{Appendix F: Sensitivity analysis}
\renewcommand {\thefigure} {F.\arabic{figure}}
\setcounter{figure}{0}

\begin{figure}
  \centering
\includegraphics[width=\textwidth]{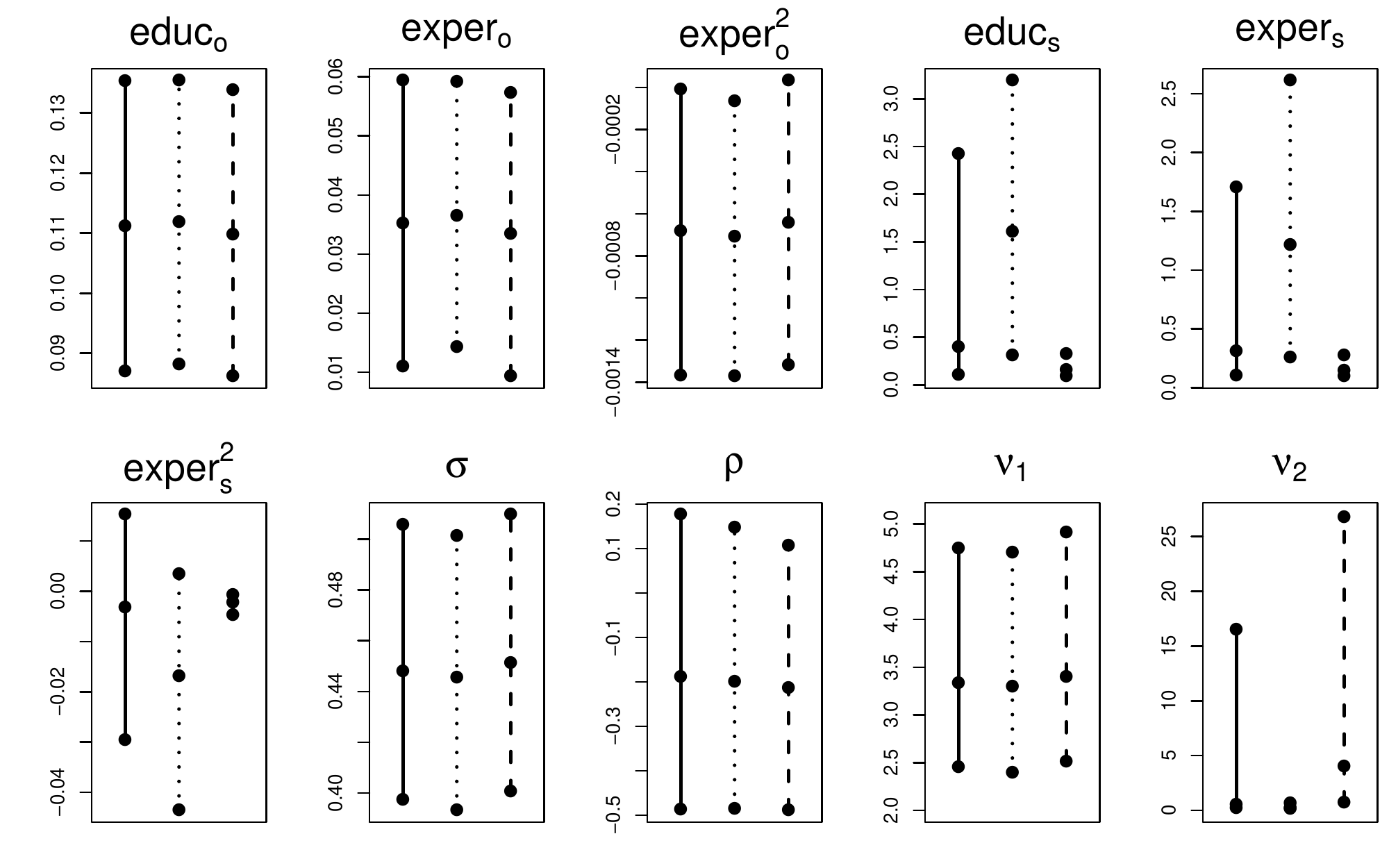}
  \caption{Wage offer function analyzed by the generalized selection-$t$ model with three different priors for $\nu$. The solid, dotted and dashed lines denote the results under the priors $\text{Gamma}(1,0.1)$, $\text{Gamma}(0.5,0.05)$ and $\text{Gamma}(1.5,0.15)$, respectively.}
  \label{fig:selection:sens} %% label for entire figure
\end{figure}

\begin{figure}
  \centering
\includegraphics[width=\textwidth]{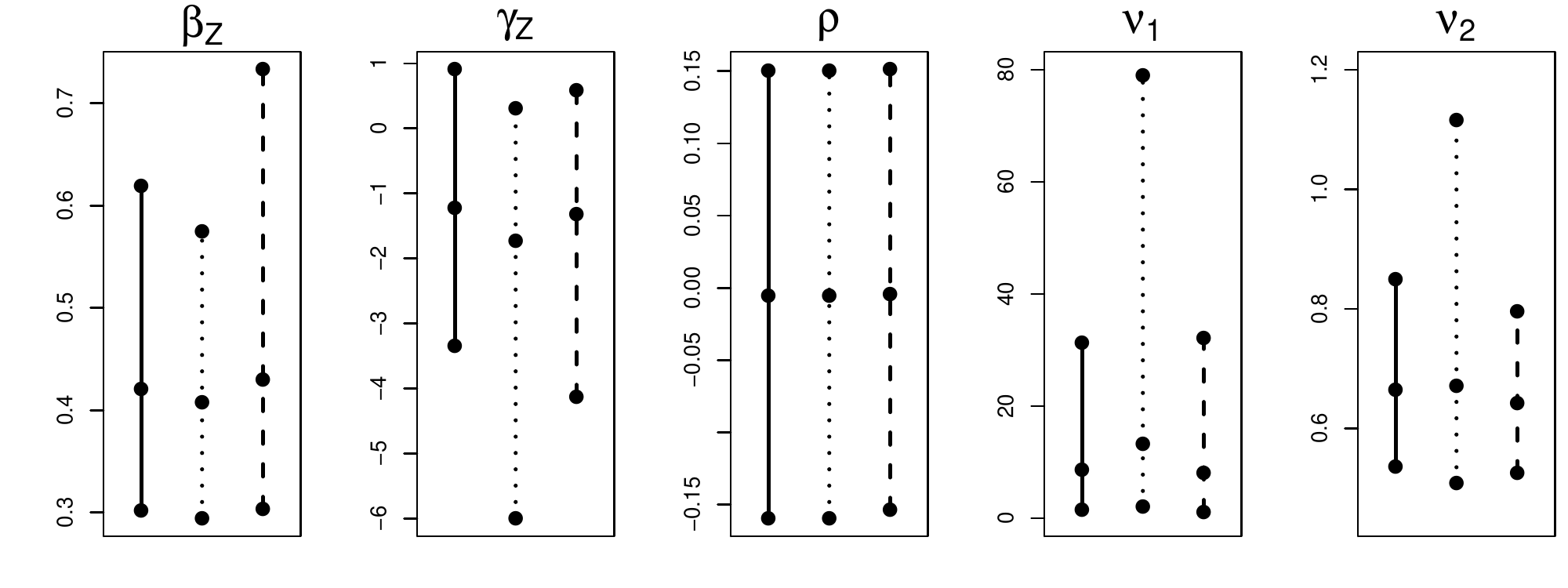}
  \caption{The flu shot experiment data analyzed by the generalized Robit model with three different priors for $\nu$. The solid, dotted and dashed lines denote the results under the priors $\text{Gamma}(1,0.1)$, $\text{Gamma}(0.5,0.05)$ and $\text{Gamma}(1.5,0.15)$, respectively.}
  \label{fig:robit:sens} %% label for entire figure
\end{figure}

\begin{figure}
  \centering
\includegraphics[width=\textwidth]{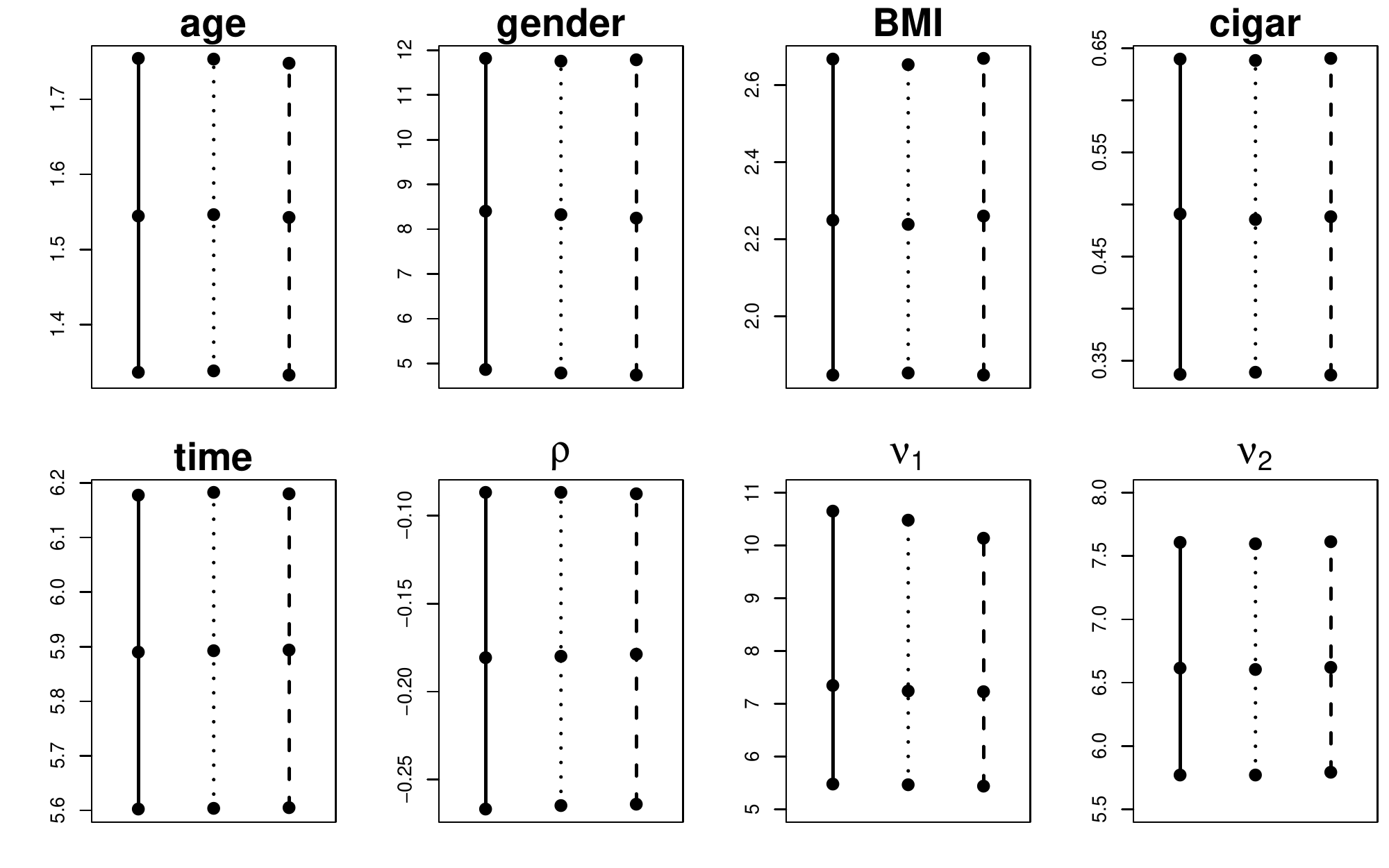}
  \caption{The Framingham study example analyzed by the generalized linear $t$ mixed-effects model with three different priors for $\nu$. The solid, dotted and dashed lines denote the results under the priors $\text{Gamma}(1,0.1)$, $\text{Gamma}(0.5,0.05)$ and $\text{Gamma}(1.5,0.15)$, respectively.}
  \label{fig:lmm:sens} %% label for entire figure
\end{figure}

To investigate the sensitivity of the results to different priors, we choose three different settings for the priors for $\bm{\nu}$ in all real examples. The priors for $\bm{\nu}$ should have wide 95\% quantile ranges, allowing for extreme heavy-tailedness, moderate heavy-tailedness, and light-tailedness. Hence, we choose the following three priors: $\text{Gamma}(1,0.1)$, $\text{Gamma}(0.5,0.05)$ and $\text{Gamma}(1.5,0.15)$, whose 95\% quantile ranges are $(0.253,36.9)$, $(0.010,50.2)$ and $(0.719,31.2)$, respectively.   Figures \ref{fig:selection:sens}--\ref{fig:lmm:sens}   show the results of the sensitivity analysis for  the generalized selection-$t$,  Robit and linear $t$ mixed-effects model, respectively. In Figures \ref{fig:robit:sens}   and  \ref{fig:lmm:sens}, the results of  the generalized Robit and linear $t$ mixed-effects model  are not sensitive to different priors of $\bm{\nu}$. In Figure \ref{fig:selection:sens}, for the generalized selection-$t$ model, the parameters in the outcome equation barely change  but the parameters in the selection equation are sensitive to different priors. However, qualitative conclusions  remain the same.


\end{document}